\documentclass[preprint,nofootinbib,superscriptaddress,amsmath,amssymb]{revtex4}

\usepackage{amsmath} 
\usepackage{graphicx} 
\usepackage{bbm}
\usepackage{etoolbox}
\preto\subequations{\ifhmode\unskip\fi}

\begin{document}

\title{BRST Quantization of Cosmological Perturbations}

\author{Cristian Armendariz-Picon}
\affiliation{Physics Department, St.~Lawrence University, Canton, NY 13617, USA.}
\email{carmendarizpicon@stlawu.edu}

\author{Gizem \c{S}eng\"or}
\affiliation{Department of Physics, Syracuse University, Syracuse, NY 13244, USA.}
\email{gizemsengor@gmail.com}

\begin{abstract}
BRST quantization is an elegant and powerful method to quantize theories with local symmetries. In this article we study the Hamiltonian BRST quantization of cosmological perturbations in a universe dominated by a scalar field, along with  the closely related  quantization method of Dirac. We describe how both formalisms apply to  perturbations in a time-dependent background, and how expectation values of gauge-invariant operators can be calculated in the in-in formalism. Our analysis focuses mostly on the free theory. By appropriate canonical transformations we simplify and diagonalize the  free Hamiltonian. BRST quantization  in  derivative gauges allows us to  dramatically simplify the structure of the propagators, whereas  Dirac quantization, which amounts to  quantization in synchronous gauge, dispenses with the need to introduce ghosts and preserves the locality of the gauge-fixed action.
\end{abstract}

\maketitle


\section{Introduction}

In order to explain the properties of the primordial fluctuations, in many models of the origin of structure one needs to quantize general relativity  coupled to a scalar field. Because the action is invariant under diffeomorphisms, the challenge one faces is the quantization of a theory with  local symmetries, very much like  those in non-Abelian gauge theories. Given the formal similarity between diffeomorphisms and the latter we shall refer to all of them as ``gauge theories."

The quantization of gauge theories is somewhat subtle, but becomes relatively straight-forward if one is interested in tree-level calculations alone: One can either fix the gauge and break the local symmetry, thus clearing the way, say,  to  canonical quantization, or one can work  with the essentially equivalent method of quantizing an appropriate set of gauge-invariant variables, a procedure also known as reduced phase space quantization.  This is the way  primordial spectra were originally calculated in the free theory \cite{Mukhanov:1990me}.  Although BRST quantization provides an elegant and powerful method  of quantization, it is not strictly necessary in those cases. 

Yet problems arise when one attempts to go beyond the free (linearized) theory. Beyond the linear order, it becomes increasingly difficult to work  with gauge-invariant variables alone (see for instance \cite{Malik:2008im,Malik:2008yp} for a Lagrangian treatment of second order perturbation theory), and beyond tree-level one needs to take into account the interactions of the ghosts associated with the gauge-fixing procedure.  From a phenomenological point of view, loop calculations in cosmological perturbation theory may not be important at this point, because observations are not sensitive enough  to these corrections (yet), but from a theoretical perspective they arguably are the distinct feature of quantum gravity, in the same way as the anomalous magnetic moment of the electron  is regarded as one of the distinct quantum signatures of QED. Among other reasons, this is why researchers  have relatively recently begun to explore loop corrections to primordial spectrum calculations  \cite{Weinberg:2005vy,loops}.  But due to the inherent complexity of these calculations, the contribution from the ghosts has been mostly ignored so far. 

The  most common method of quantizing a gauge theory is that of Faddeev and Popov (and also DeWitt.) This method is inherently linked to the choice of gauge-fixing ``conditions." In the case of cosmological perturbations, these can be taken to be four functions $f^\mu$ of the inflaton and metric perturbations $\delta\varphi,\delta h_{\mu\nu}$ that are \emph{not} invariant under the four independent diffeomorphisms with infinitesimal parameters $\xi^\nu$. Once these conditions have been chosen, the action of the theory needs to be supplemented with appropriate ghost terms,
\begin{equation}\label{eq:FP0}
	S_\mathrm{ghost} =\int d^4 x \, d^4 y \, \,
	\bar \eta_\mu  (x)
		\frac{\delta f^\mu (x)}{\delta \xi^\nu(y)}
	\eta^\nu(y),
\end{equation}
which typically couple the fermionic  Faddeev-Popov ghosts  $\bar\eta_\mu$ and $\eta^\nu$ to the metric and inflaton perturbations. The Faddeev-Popov method works well in renormalizable gauge theories, provided that the gauge-fixing functionals are linear in the fields. But if the gauge-fixing functionals are non-linear, or the theory is non-renormalizable,  the Lagrangian needs to be supplemented with terms that are not just quadratic in the ghosts. Since these factors are absent in the Faddeev-Popov prescription (\ref{eq:FP0}), the method fails.  

A very general and powerful  quantization method that avoids these problems, and reduces to the one  of Faddeev and Popov in appropriate cases, is that of Becchi, Rouet, and Stora, and Tyutin (BRST) \cite{BRST}.  The BRST method not only justifies Faddeev and Popov, but  also endows it with a geometric interpretation. In its Hamiltonian formulation, BRST quantization manifestly results in a unitary theory, and it also manifestly preserves a global supersymmetry known as BRST symmetry, even after gauge-fixing.  This quantization method has been successfully applied in many different contexts, ranging from electrodynamics to string theory, and  is arguably  the best way to quantize a theory with local symmetries.

There are many important reasons for pursuing BRST quantization in the context of cosmological perturbations. As we already mentioned,   the method of Faddeev and Popov fails when applied to gravity. In addition,   BRST quantization  provides us with an enormous freedom to choose gauge-fixing terms. While in field theories in Minkowski spacetime the demand of Lorentz invariance and renormalizability severely restricts the possible gauge-fixing choices, in cosmological perturbation theory the smaller degree of symmetry allows for a much wider set of gauge conditions that have remained essentially unexplored so far.  Calculations in cosmological perturbation theory are notoriously involved, and an eventual simplification of the propagators and the structure of the ghost interactions facilitated by appropriate generalized gauge choices may render loop calculations in the BRST method much  more manageable. The BRST global symmetry preserved even after  gauge fixing may  also place interesting constraints on the structure of the theory that describes the cosmological perturbations and its implications too (see \cite{Binosi:2015obq} for a  discussion in the context of the  antifield formalism.)  On a related topic, we should also note that whereas typical gauge choices in cosmological perturbation theory break locality,   BRST quantization allows for gauge-fixing conditions that manifestly preserve the latter. Locality (analyticity in the spatial momenta) was for instance an important ingredient in the derivation of consistency relations between cosmological correlators derived in \cite{Berezhiani:2013ewa,Armendariz-Picon:2014xda}.

In this article we study the BRST quantization of cosmological perturbations. We mostly concentrate on the general formalism and illustrate many of the results in the free theory. We approach the quantization from the Hamiltonian perspective, which is manifestly unitary, and makes the role of boundary conditions more explicit.  In a cosmological background, the split into space and time required by the Hamiltonian formulation does not conflict with any spacetime isometry, and thus does not pose any immediate significant drawback.  If it exists, the Lagrangian formulation can be recovered from the Hamiltonian one by integration over the canonical momenta as usual.  For lack of space, however, we do not discuss  the  powerful Lagrangian  antifield formalism of Batalin and Vilkovisky, which we hope to explore in future work. 

 We have tried to make the article relatively self-contained, which is why we quote  main results in BRST quantization, at the expense of making the manuscript  longer than strictly necessary. Our presentation mostly follows the excellent monograph  by Henneaux and Teiltelboim \cite{Henneaux:1992ig}, which the reader may want to consult for further background and details. To our knowledge, our work is  the first to focus on the Hamiltonian BRST quantization of cosmological perturbations, although Barvinsky has discussed the BRST formalism in the context of a cosmological density matrix \cite{Barvinsky:2013nca}, and Binosi and Quadri have used the  antifield formalism to reproduce some of the consistency relations satisfied by cosmological correlators \cite{Binosi:2015obq}. Other authors have analyzed somewhat related issues, mostly  within the path integral quantization of   cosmological perturbations \cite{Anderegg:1994xq,Prokopec:2010be}, or within the loop quantum cosmology program \cite{Ashtekar:2011ni}. The Hamiltonian of cosmological perturbations    was  calculated  to quadratic order in reduced phase space in \cite{Langlois:1994ec}, and  to cubic order  only in spatially flat gauge  \cite{Nandi:2015ogk}, due to the above-mentioned complexity of the gauge-invariant formalism at higher orders. An intriguing approach that aims at formulating the Hamiltonian of the theory directly in terms of gauge-invariant variables (to all orders) is discussed in \cite{Giesel:2007wi}. The latter introduces a pressureless fluid parameterized by four spacetime scalars, which are used to define gauge-invariant observables and a gauge-invariant Hamiltonian by deparameterization. Such approach is in many ways complementary to the one of the BRST formalism, which is built around gauge-variant fields, and ultimately relies on a choice of gauge. Both share the property that a pair of additional  fields is introduced for each of the four constraints of the theory,  but whereas BRST-invariance guarantees that the ghosts do not change the gauge-invariant content of the theory, the dust fields of \cite{Giesel:2007wi}  do seem to ultimately survive as additional degrees of freedom in the system.

\section{Action}

Our main goal is the quantization of cosmological perturbations in a spatially flat universe dominated by a canonical scalar field. This is for instance what is needed to calculate primordial perturbation spectra in conventional inflationary models, although our results do not really depend on any particular scalar field background, as long as the latter is homogeneous and time-dependent. 

We begin with the action of general relativity minimally coupled to a scalar field $\varphi$ in Hamiltonian form.
As  is well known, the Hamiltonian action takes its most natural form in the ADM formulation \cite{Arnowitt:1962hi}, in which the metric components are written as 
\begin{equation}\label{eq:ADM}
	ds^2=-(\lambda^N)^2 dt^2+h_{ij}(dx^i+\lambda^i dt)(dx^j+\lambda^j dt),
\end{equation}
where $\lambda^N$ is the lapse function and $\lambda^i$ the shift vector.  With this  choice of variables the action of the theory becomes
\begin{equation}\label{eq:Hamiltonian S}
	S=\int dt\, \int d^3x \, \left(\pi^{ij}\, \dot h_{ij} +\pi_\varphi \, \dot \varphi\right) -H,        
\end{equation}
where the Hamiltonian  $H$  is linear in  $\lambda^N$ and $\lambda^i$,
\begin{equation}\label{eq:H}
	H=\int d^3x\,\left[\lambda^N \, G_N +\lambda^i\, G_i\right],
\end{equation} 
with coefficients given by
\begin{subequations}\label{eq:secondary}
\begin{align}
	G_N &\equiv 
	\frac{2}{M^2}\frac{1}{\sqrt{h}} \left(\pi_{ij}\pi^{ij} -\frac{\pi^2}{2}\right)-\frac{M^2}{2}\sqrt{h}R^{(3)}
	+\frac{\pi_\varphi^2}{2\sqrt{h}}
	+\frac{\sqrt{h}}{2}h^{ij}\partial_i\varphi \partial_j\varphi
	+\sqrt{h}\,V(\varphi), \label{eq:Hamiltonian C}
	\\
	G_i&\equiv 
	-2\sqrt{h}\nabla_j\left(\frac{\pi_i{}^j}{\sqrt{h}} \right)+\pi_\varphi \partial_i \varphi
	. \label{eq:momentum C}
\end{align}
\end{subequations}
To arrive at these expressions  we have discarded a surface term at the spatial boundary. Indices are raised and lowered with the spatial metric $h_{ij}$, and $M$ is the reduced Planck mass.  

Because the action does not contain derivatives of the lapse function $\lambda^N$ and shift vector $\lambda^i$,  their conjugate momenta vanish, $b_N\equiv 0, b_i\equiv 0$. 
In Dirac's analysis, these would be interpreted as primary constraints.  But in the end, this interpretation mostly leads to unnecessary complications. In the Hamiltonian formulation it is much simpler  to think of $\lambda^N$ and $\lambda^i$ as Lagrange multipliers, and restrict phase space to the appropriately constrained  canonical pairs
$\{(h_{ij},\pi^{ij}), (\varphi,\pi_\varphi)\}$. 

Variation of the action  (\ref{eq:Hamiltonian S}) with respect to  the $\lambda^N$ and $\lambda^i$  yields the secondary constraints $G_N=G_i=0$, which is why we loosely refer to $G_N$ and $G_i$ as ``the constraints." These constraints define gauge transformations on phase space functions $F$ through  their Poisson brackets,
$
	\Delta_a F\equiv\{F,G_a\}. 
$
In this way, the  $G_i$ generate spatial diffeomorphisms.  The constraint $G_N$ generates diffeomorphisms along the normal to the equal-time hypersurfaces only after the equations of motion are imposed on the Poisson bracket.  For our purposes, what matters most is that the constraints are first class, and that they define an  open algebra  with field-dependent structure constants (the Dirac algebra.) In practice this means that their Poisson brackets are just proportional to the constraints themselves \cite{Isham:1992ms}, 
\begin{subequations}\label{eq:algebra}
\begin{align}
	\left\{G_N(\vec{x}),G_N(\vec{y})\right\}&=
	\frac{\partial\delta(\vec{y}-\vec{x})}{\partial x^i}h^{ij}(\vec{x}) G_j(\vec{x})-\frac{\partial\delta(\vec{x}-\vec{y})}{\partial y^i}h^{ij}(\vec{y})G_j(\vec{y}), \label{eq:GN GN}
	\\
	\left\{G_N(\vec{x}),G_i(\vec{y})\right\}&=\frac{\partial\delta(\vec{x}-\vec{y})}{\partial x^i}G_N(\vec{y}),
	\\
	\left\{G_i(\vec{x}),G_j(\vec{y})\right\}&=\frac{\partial \delta(\vec{x}-\vec{y})}{\partial x^i}G_j(\vec{x})-\frac{\partial \delta(\vec{y}-\vec{x})}{\partial y^j} G_i(\vec{y}),
\end{align}
\end{subequations}
with coefficients that,  in the case of equation (\ref{eq:GN GN}), depend on the inverse spatial metric $h^{ij}$. Because of this dependence, the commutator of two gauge transformations is another gauge transformation only after the constraints are imposed on the result. In that sense, the algebra of the constraints only closes on-shell, which is why one refers to it as  an open algebra.  For simplicity, we shall nevertheless refer to the constraints as the generators of diffeomorphisms.  The Hamiltonian (\ref{eq:H})  itself  is thus a linear combination of the  four secondary constraints, and  vanishes identically on shell.  Because its  Poisson brackets with the secondary constraints vanish weakly,\footnote{A function of the canonical variables is said to vanish weakly, when it vanishes after the application of the constraints.} no further constraints appear in the theory. Whereas  in the Hamiltonian formalism the algebra of constraints is open, in the Lagrangian formalism the generators of diffeomorphisms along the four spacetime coordinates define a closed algebra: The commutator of two diffeomorphisms with  infinitesimal parameters $\xi_1^\mu$ and $\xi_2^\mu$ is  a diffeomorphism with infinitesimal parameters  $\xi_1^\nu\partial_\nu\xi_2^\mu-\xi_2^\nu\partial_\nu \xi_1^\mu$.

\subsection{Perturbations}

We are actually interested in quantizing the perturbations around a cosmological background. Therefore, we split the ADM  variables into  background plus perturbations,
\begin{subequations}\label{eq:background+perturbations}
\begin{align}
	\varphi&=\bar{\varphi}(t)+\delta\varphi,
	&\pi_\varphi&=\bar\pi_\varphi+\delta \pi_\varphi,\\
	h_{ij}&=\bar h_{ij}+\delta h_{ij},
	&\pi^{ij}&=\bar\pi^{ij}+\delta\pi^{ij},\\
	\lambda^N&=\bar{\lambda}^N+\delta\lambda^N,\\
 	\lambda^i&=\bar{\lambda}^i+\delta\lambda^i,
\end{align}
\end{subequations}
where the background quantities are
\begin{equation}\label{eq:background}
	\bar{\lambda}^N=a,
	\quad
	\bar \lambda^i=0,
	\quad
	\bar{h}_{ij}=a^2\delta_{ij},
	\quad
	 \bar{\pi}^{ij}=-M^2 \mathcal{H} \, \delta^{ij},
	 \quad
	  \bar{\pi}_\varphi=a^2 \dot{\bar{\varphi}},
\end{equation}
and $a$ is the scale factor. Note that our choice of background lapse function, $\bar{\lambda}^N\equiv a$, implies that the time coordinate $t$ is actually conformal time, rather than the conventional cosmic time. In addition, our definition of the metric perturbations $\delta h_{ij}$ does not separate a factor of $a^2$ from the perturbations. A dot denotes a derivative with respect to coordinate time (conformal time), and
$\mathcal{H}=\dot{a}/a$ stands for the comoving Hubble scale. The background quantities obey the equations of motion
\begin{subequations}\label{eq:background motion}
\begin{align}
	\ddot{\bar\varphi}+2\mathcal{H}\dot{\bar\varphi}+a^2 \bar{V}_{,\varphi}&=0,\\
	2\dot{\mathcal{H}}+\mathcal{H}^2+\frac{1}{M^2}\left(\frac{\dot{\bar\varphi}^2}{2}-a^2\bar{V}\right)&=0,
\end{align}
as well as the single constraint
\begin{equation}
\mathcal{H}^2-\frac{1}{3M^2}\left(\frac{\dot{\bar\varphi}^2}{2}+a^2\bar{V}\right)=0.
\end{equation}
\end{subequations}
We shall often use these relations to  simplify some of the resulting expressions throughout this work. 

To simplify the notation, a transition to DeWitt notation proves to be advantageous.  In this notation equations (\ref{eq:background+perturbations}) become
\begin{equation}\label{eq:canonical T}
	 q^i\equiv \bar{q}^i+\delta q^i,
	 \quad 
	p_i\equiv \bar{p}_i+ \delta p_i,
	\quad
	 \lambda^a\equiv \bar\lambda^a+\delta\lambda^a,
\end{equation}
where the $q^i$ stand for the canonical variables $ h_{ij}(\vec{x})$ and $\varphi(\vec{x})$, the $p_i$ for their conjugate momenta, and the  $\lambda^a$ for the Lagrange multipliers $\lambda^N(\vec{x})$ and $\lambda^i(\vec{x})$. In particular, in DeWitt notation indices like $i$ and $a$ stand both for field indices and spatial coordinates.

To obtain the action for the perturbations we simply substitute the expansion (\ref{eq:background+perturbations}) into  the action (\ref{eq:Hamiltonian S}). The Hamiltonian action for the perturbations then  becomes
\begin{equation}\label{eq:S pert I}
\delta S=\int dt \left[\delta p_i \delta\dot q^i-\delta H\right],
\end{equation}
in which the  Hamiltonian for the perturbations is obtained from the original  one by removing linear terms in the canonical perturbation variables. In the case at hand, the Hamiltonian of the full theory  is just a linear combination of the constraints, so the  Hamiltonian of the perturbations becomes 
\begin{equation}\label{eq:delta H}
	\delta H=-\dot{\bar q}^i \delta p_i+\dot{\bar p}_i\delta q^i +(\bar\lambda^a+\delta\lambda^a)G_a(\bar{p}_i+\delta p_i,\bar{q}^i+\delta q^i).
\end{equation}
The first two terms in the Hamiltonian (\ref{eq:delta H}) are just the result of a time-dependent canonical transformation applied to a vanishing Hamiltonian.  Of course, although we have changed variables, the original constraints are still satisfied. In particular,  variation with respect to $\delta\lambda^a$ yields again 
\begin{equation}\label{eq:constraints pert}
 \delta G_a(\delta p_i, \delta q^i,t)\equiv G_a(\bar{p}_i+\delta p_i,\bar{q}^i+\delta q^i)=0.
 \end{equation}
Our notation aims to emphasize that the $\delta G_a$ are functions of the perturbations and time.  They have the same value as the constraints $G_a$, but  their functional forms differ. 

 Enforcing the constraints in equation (\ref{eq:delta H}) returns a Hamiltonian linear in the perturbation variables that is not particularly useful for cosmological perturbation theory calculations. Instead, it shall  prove to be more useful to separate only the contribution of the Lagrange multipliers $\delta\lambda^a$ to the Hamiltonian. We thus write
 \begin{equation}\label{eq:H HD}
 	\delta H=\delta H_D+\delta\lambda^a \delta G_a,
 \end{equation}
where the ``first-class Hamiltonian" is 
\begin{equation}\label{eq:H Dirac}
	\delta H_D=-\dot{\bar q}^i\delta p_i+\dot{\bar p}_i\delta q^i +\bar\lambda^a \delta G_a.
\end{equation}
Note that because the background satisfies the equations of motion, the linear terms in this expression cancel. As we show below,  under the time evolution defined by $\delta H_D$  the constraints are also preserved weakly, or, alternatively, the Hamiltonian $\delta H_D$ is weakly gauge-invariant (hence its name.) The first-class Hamiltonian plays an important role in the BRST theory.

The time derivative of any function of the perturbations $\delta F$ follows from its Poisson brackets with the Hamiltonian,
\begin{equation}\label{eq:delta t pert}
	\frac{d\delta F}{dt}=\{\delta F,\delta H\}+\frac{\partial \delta F}{\partial t}. 
\end{equation} 
We include a partial derivative with respect to time to capture a possible explicit time dependence of $\delta F$, say, due to its dependence on  background quantities. Nevertheless, if $\delta F$ is of the form $\delta F(\delta q^i,\delta p_i)=F(\bar{q}^i+\delta q^i,\bar{p}_i+\delta\bar{p}_i),$ its time evolution obeys
\begin{equation}\label{eq:dt delta F}
	\frac{d\delta F}{dt}=\{F,H\},
\end{equation}
with the understanding that after calculation of the Poisson bracket on the right hand side in the original unperturbed variables, the substitution (\ref{eq:canonical T}) is supposed to be made.  The perturbation Hamiltonian (\ref{eq:delta H}) weakly equals $-\dot{\bar{q}}^i \delta p_i+\dot{\bar{p}}_i\delta q^i$, which is generically   not weakly conserved.  On the other hand, the original Hamiltonian $H$, understood as function of the perturbations does remain constant. In fact, it identically vanishes.

\subsection*{Analysis of the Constraints}
Variation of the action with respect to the perturbed Lagrange multipliers yields equations (\ref{eq:constraints pert}). These are simply the original constraints expressed in terms of the perturbations.  Because the transition to the perturbations is a canonical transformation, the Poisson brackets of any phase space function is preserved under such a change of variables. In particular, it follows that 
\begin{equation}
\{\delta q^i, \delta G_a\}=\{q^i, G_a\}, \quad
\{\delta p_i, \delta G_a\}=\{p_i, G_a\},
\end{equation}
where, again, on the expressions on the right hand sides $q^i$ and $p_i$ are to be expanded as in equations (\ref{eq:canonical T}) after calculation of the bracket. Therefore, the constraints act on the perturbations as they did on the original variables. Since, as we mentioned, the constraints $G_a$ generate diffeomorphisms when they act on $p_i$ and $q^i$, so do the $\delta G_a$ when they act on $\delta p_i$ and $\delta q^i$.  It also    follows  immediately that
$
\{\delta G_a,\delta G_b\}=\{G_a,G_b\}(\bar{p}_i+\delta p_i,\bar q^i+\delta q^i).
$
Because the Poisson bracket $\{G_a,G_b\}$ is linear in the constraints, and because of  equation (\ref{eq:constraints pert}), the algebra of perturbed constraints remains the same. In particular, the set of constraints $\delta G_a=0$ is first class. In DeWitt notation we write the algebra of constraints as
\begin{equation}\label{eq:structure}
\{\delta G_a,\delta G_b\}=C_{ab}\,{}^c\delta G_c,
\end{equation} 
and call the field-dependent coefficients $C_{ab}{}^c$ the ``structure constants" of the theory. But a bit of care must be taken because the constraints $\delta G_a$ now depend explicitly on time, so the Poisson brackets alone do not  fully determine their time  evolution. Nevertheless, using equation (\ref{eq:dt delta F}) we find
\begin{equation}\label{eq:dt delta C}
\frac{d \delta G_a}{dt}=(\bar\lambda^b+\delta \lambda^b)\{\delta G_a,\delta G_b\},
\end{equation}
which indeed vanishes weakly because of equation (\ref{eq:structure}). As a result, no equation in the theory fixes the value of the Lagrange multipliers $\delta \lambda^a$, which remain undetermined by the equations of motion, as  in the background-independent theory. 

\subsection*{Gauge Symmetry}

By definition, in the Hamiltonian formulation the gauge transformations generated by the constraints  act only on the canonical variables $\delta q^i$ and $\delta p_i$.  They capture the redundancy in the description of the state of the system at any particular time.  It is also interesting to elucidate to what extent the first-order action (\ref{eq:S pert I}) is symmetric under diffeomorphisms, and how the latter act on the Lagrange multipliers. By direct substitution, the reader can check that, indeed, the transformations
\begin{subequations}\label{eq:Delta delta}
\begin{align}
	\Delta \delta q^i&=\xi^a \{\delta q^i,\delta G_a\}, \\
	\Delta \delta p_i&=\xi^a\{\delta p_i, \delta G_a\},\\
	\Delta \delta \lambda^a&=\dot{\xi}^a
	+\xi^b(\bar{\lambda}^c+\delta\lambda^c) C_{bc}{}^a,
	\label{eq:Delta delta lambda}
\end{align}
\end{subequations}
leave the Hamiltonian action for the perturbations invariant for arbitrary $\xi^a(t)$, provided that the latter vanish at the time boundary.  The structure of the transformation laws  (\ref{eq:Delta delta}) follows from the properties of the constraints and the first-class Hamiltonian $\delta H_D$. Indeed, using  equations (\ref{eq:H HD}) and  (\ref{eq:dt delta C}) we find
\begin{equation}\label{eq:first class}
	\{\delta G_a,\delta H_D\}+\frac{\partial \delta G_a}{\partial t}=\bar\lambda^b C_{ab}{}^c \delta G_c.
\end{equation}
The term $\xi^b\bar\lambda^c C_{bc}{}^a$ in equation (\ref{eq:Delta delta lambda}) is what one expects in a theory in which the first-class Hamiltonian satisfies equation (\ref{eq:first class}), whereas the term $\xi^b\delta\lambda^c C_{bc}{}^a$  is what one should have in a gauge theory with structure constants $C_{ab}{}^c$. 

\section{BRST Symmetry}

In the classical theory, the existence of local symmetries leads to equations of motion that do not admit unique solutions with the prescribed boundary conditions, because the Lagrange multipliers remain arbitrary. There are mainly two ways of dealing with such ambiguities: One can either work with gauge-invariant variables alone, as in the reduced phase space method, or one can simply fix the gauge. Working with gauge-invariant variables becomes increasingly complex and cumbersome at higher orders in perturbation theory, so we choose here the much simpler alternative of fixing the gauge.  One of the key aspects of gauge theories is that, even after gauge fixing, they retain a global symmetry known as BRST symmetry, so named after Becchi, Rouet, Stora, and Tyutin \cite{BRST}. We introduce this symmetry next, mostly by following the presentation of \cite{Henneaux:1992ig}.  An alternative  way to reach some of the results here consists of identifying  the BRST symmetry directly in the background-independent theory, and then making a transition to the theory of the perturbations by canonical transformation. 

\subsection{Ghosts and Antighosts}

 To bring the BRST symmetry to light, we introduce for each  constraint $\delta G_a=0$ a  real Grassmann  variable  $\delta\eta^a$ known as the ``ghost", and its purely imaginary  conjugate momentum $\delta \mathcal{P}_a$. The ghosts $\delta \eta^a$ carry ghost number one, and their conjugate momenta carry ghost number minus one.   These variables commute with all bosonic fields and anticommute with all fermionic fields.\footnote{The previous statements  imply for instance that a term $\delta\dot\eta^a \delta\mathcal{P}_a $ in the first-order ghost action is real and has ghost number zero.} In particular, ghosts and their momenta obey the  Poisson bracket relations
\begin{equation}
	\{\delta \mathcal{P}_a,\delta \eta^b\}\equiv
	-\delta_a{}^b. 
\end{equation} 
From now on, the brackets $\{\cdot,\cdot\}$ refer to the graded bracket of any two functions defined on the extended phase space spanned by the canonical pairs $(\delta p_i,\delta q^i)$ and $(\delta\mathcal{P}_a, \delta\eta^a)$. We summarize the properties of the graded bracket in Appendix \ref{sec:Graded Commutator}.   

We could also think of these ghosts as  perturbations around a background with vanishing ghosts. The transformation properties of the former under the isometries of the background are those expected from their indices. Namely, $\delta\eta^N$ and $\delta\eta^i$ respectively transform as scalars under translations, and as a scalar and a three-vector under rotations.

For some purposes, it shall prove to be convenient to introduce yet another set of  ghosts associated with the primary constraints $\delta b_N\equiv\delta b_i\equiv 0$ that we opted to bypass for simplicity earlier on. We shall denote these (Grassmann) ghosts by  $\delta C_a$, and their conjugate momenta by $\delta\rho^a$. The  $\delta C_a$ are real and carry ghost number minus one; they  are thus known as ``antighosts." Their conjugate momenta are purely imaginary and carry ghost number one. These variables accordingly obey the Poisson bracket relations
\begin{equation}
	\{\delta \mathcal{\rho}^a,\delta C_b\}=-\delta^a{}_b.
 \end{equation}
  
In the work of Faddeev, the ghosts were simply introduced to represent the Faddeev-Popov determinant $\det \{f^a,\delta G_b\}$ as a quadratic functional integral in phase space \cite{Faddeev:1969su}. Such a determinant renders the  expectation value of an observable independent of the gauge-fixing conditions $f^a=0$, so in this approach the ghosts were simply regarded as a necessary by-product  of any gauge-fixing procedure. But the   ghosts $\delta\eta^a$ and their conjugates $\delta\mathcal{P}_a$ also have a geometrical interpretation that only  emerged after the discovery of the BRST symmetry. In the BRST formalism one identifies the observables of the gauge theory with the cohomology of an appropriately defined BRST differential. In a gauge theory, the observables consist of those functions defined on the constraint surface $\delta G_a=0$ that  remain invariant under the transformations generated by the  constraints. On one hand, in order to identify functions on the constraint surface with the cohomology of a differential operator, one needs to introduce as many Grassmann variables $\delta\mathcal{P}_a$ as there are conditions  defining the constraint surface. On the other hand,  in order to identify the cohomology of a differential operator with those functions invariant under the constraints, one needs to introduce as many differential forms $\delta\eta^a$ as there are gauge generators. These forms can be thought of as the  duals to the tangent vector fields  defined by the generators, and, along with their exterior product,  they define a Grassmann algebra.  In a theory with first-class constraints the functions that define the constraint surface are  the same as the gauge generators, which allows one to identify the $\delta\mathcal{P}_a$ as the canonical conjugates of the $\delta\eta^a$.

\subsection{BRST Transformations}

The main idea behind BRST quantization is the replacement of the local symmetry under gauge transformations by a nilpotent  global symmetry generated by what is known as the BRST charge.   This  BRST charge is defined to be a real, Grassmann-odd, ghost number one and   nilpotent function in the  extended phase space introduced above. For many purposes, it is convenient to expand this BRST charge in the antighost number $p$, that is, in powers of the fields $\delta\mathcal{P}_a$,
\begin{equation}\label{eq:omega}
	\delta \Omega=\delta \eta^a \delta G_a-\frac{1}{2}\delta \eta^b \delta \eta^c C_{cb}{}^a \delta \mathcal{P}_a+\delta \Omega_{p\geq 2},
\end{equation}
where the structure constants $C_{ab}{}^c$ are those of equation (\ref{eq:algebra}), and $\delta \Omega_{p\geq 2}$ is of antighost number $p\geq 2$.  The latter vanishes for closed algebras, and can be otherwise determined recursively by demanding that the BRST charge be of ghost number one and obey the nilpotence condition
\begin{equation}\label{eq:nilpotence}
	\{\delta\Omega,\delta\Omega\}=0.
\end{equation}
The BRST charge $\delta\Omega$ captures the gauge symmetries of a theory, and not the variables we choose to describe it. In particular, we could have arrived at equation (\ref{eq:omega}) by identifying the gauge symmetries and their algebra in the background-independent theory, and then performing a canonical transformation to the perturbations around our cosmological background.  Such a procedure manifestly underscores that $\delta\Omega$ has the structure
\begin{equation}\label{eq:omega structure}
	\delta\Omega(\delta p_i,\delta q^i; \delta\mathcal{P}_a,\delta \eta^a)=\Omega (\bar{p}_i+\delta p_i,\bar{q}^i+\delta q^i ;\delta\mathcal{P}_a,\delta \eta^a),
\end{equation}
where $\Omega$ is the BRST charge of the background-independent theory. As we show below, this structure guarantees that the BRST action for the perturbations remains BRST-symmetric.

The BRST charge defines a transformation $s$ on any function of the extended phase space through its graded bracket,
\begin{equation}\label{eq:BRST T}
	s \,  \delta F=\{\delta F,\delta\Omega\}.
\end{equation}
Its action on the canonical variables contains a gauge transformation with the ghosts as gauge parameters, plus terms of antighost number $p\geq 1$ (denoted by dots), 
\begin{subequations}
\begin{align}
	s \, \delta q^i & =\delta \eta^a \{ \delta q^i,\delta G_a\}+\cdots
	, 
	&s \,   \delta p_i &=\delta \eta^a \{\delta  p_i,\delta G_a\}+\cdots
	,\\ 
	s \, \eta^a&=\frac{1}{2}\delta \eta^b \delta\eta^c C_{cb}{}^a+\cdots
	,
	&s\, \delta\mathcal{P}_a&=-\delta G_a+\cdots.
\end{align}
\end{subequations}
These  BRST transformations are  nilpotent, $s^2=0$, because of equations (\ref{eq:nilpotence}) and (\ref{eq:Jacobi}). If under  translations and rotations the ghosts transform as indicated by their indices, we expect the BRST charge to be a scalar.

In a gauge theory, the observables consist of the set of all gauge-invariant functions on the constraint surface. One of the fundamental results in BRST theory is that this set is in one-to-one correspondence with  the cohomology of the BRST operator $s$ at  ghost number zero.
The cohomology  consists of all those functions in the extended phase space  of ghost number zero that are BRST-closed ($s \delta\mathcal{O}=0$) modulo those functions that are BRST-exact ($\delta \mathcal{O}=s\delta \mathcal{Q}$.)  If a member of the zero ghost cohomology $\delta\mathcal{O}_\mathrm{BRS}$  satisfies
\begin{equation}
	(\delta\mathcal{O}_\mathrm{BRS})|_{p=0}=\delta \mathcal{O}, 
\end{equation}
where $\delta\mathcal{O}=\delta\mathcal{O}(p_i, q^i)$ is an observable, and $|_{p=0}$ is the restriction  to those terms of antighost number zero, one speaks of $\delta\mathcal{O}_\mathrm{BRS}$ as a ``BRST-invariant extension" of the observable $\delta \mathcal{O}$. In theories with Abelian constraints, such as the linearized theory of cosmological perturbations, gauge-invariant operators are automatically BRST-invariant, and one can construct further BRST-invariant extensions by multiplication with  a BRST-invariant extension of the identity operator (see below.) In theories with a non-minimal sector, the BRST charge is
\begin{equation}\label{eq:omega nm}
	\delta\Omega^\mathrm{nm}=\delta \Omega-i \delta\rho^a \delta b_a,
\end{equation}
where $\delta\Omega$ is the same as in equation (\ref{eq:omega}). The extra contribution can be thought to originate from the additional constraints $\delta b_a=0$ that appear when one regards the Lagrange multipliers as phase space variables. The only difference is that  we think of the $\delta\rho^a$ as canonical momenta, rather than configuration variables.

\subsection{Time Evolution}

Because of the explicit time dependence, neither $\delta H$  nor $\delta H_D$ in equation (\ref{eq:H HD}) are gauge-invariant, so we cannot define their BRST-invariant extensions in the previous sense.  Instead, we define a ``BRST extension" of the Hamiltonian to be any  function $\delta H_\mathrm{BRS}$ of ghost number zero that generates  a time evolution under which the BRST charge is conserved, and whose zero antighost component agrees with the first-class Hamiltonian, 
\begin{subequations}\label{eq:BRST extension}
\begin{align}
	\{\delta\Omega,\delta H_\mathrm{BRS}\}+\frac{\partial\delta \Omega}{\partial t}&=0,
	\label{eq:H extension}\\
	(\delta H_\mathrm{BRS})_{p=0}&=\delta H_D. 
\end{align}
\end{subequations}
Because the BRST transformation is nilpotent, the BRST extension of the Hamiltonian is not unique. Indeed, if $\delta H_\mathrm{BRS}$ satisfies equation (\ref{eq:H extension}), so does 
\begin{equation}\label{eq:H K}
	\delta H_K\equiv \delta H_\mathrm{BRS}+\{\delta K,\delta\Omega\}
\end{equation}
for any  Grassmann-odd extended phase space function $\delta K$ of ghost number  minus one.  In addition, up to terms proportional to the constraints, the $p=0$ terms of both $\delta H_K$ and $\delta H_\mathrm{BRS}$ agree. 

The function $\delta K$ is known as the gauge-fixing fermion, and the Hamiltonian (\ref{eq:H K}) as the gauge-fixed Hamiltonian. This gauge-fixed Hamiltonian determines the time evolution of any function $\delta F$ in the extended phase space through its Poisson bracket as usual,
\begin{equation}\label{eq:dt BRST}
	\frac{d\delta F}{dt}=\{\delta F,\delta H_K\}+\frac{\partial\delta F}{\partial t},
\end{equation}
By definition, under such  evolution the BRST charge is conserved.   The conservation of  $\delta\Omega$ in the theory of the perturbations essentially follows from the conservation of the background-independent  $\Omega$  under a time evolution generated by an identically vanishing Hamiltonian. 

\subsection{BRST symmetry}

Because of the explicit time dependence, the gauge-fixed Hamiltonian $\delta H_K$  is not BRST-invariant. Nevertheless, by definition, the transformation induced by $\delta\Omega$ is a global symmetry, in the sense of the conservation law (\ref{eq:H extension}).  In fact, the BRST equations of motion in the minimal sector can be derived from the variational principle
\begin{equation}\label{eq:delta S BRST}
	\delta S_K=\int dt\left[\delta\dot{q}^i \delta p_i+\delta\dot{\eta}^a\delta\mathcal{P}_a-\delta H_K\right].
\end{equation}
By direct substitution, one can check that, up to boundary terms, this action is invariant under the infinitesimal BRST transformation  $\Delta_\theta F\equiv \{F,\theta \delta\,\Omega\}$, where $\theta$ is a constant  Grassmann-odd parameter. Note that the variables $\delta\lambda^a$  do not appear in the action (\ref{eq:delta S BRST}). They show up in the non-minimal sector of the theory, whose dynamics is generated by the action
\begin{equation}\label{eq:delta S BRST non-min}
	\delta S^{\mathrm{nm}}_K=\int dt\left[\delta\dot{q}^i \delta p_i+\delta\dot{\eta}^a\delta\mathcal{P}_a
	+\delta\dot\lambda^a \delta b_a+\delta\dot{C}_a\delta\rho^a
	-\delta H_K\right].
\end{equation}
In this case, the variables $\delta\lambda^a$ and their conjugate momenta $\delta b_a$ belong to the phase space of the theory.  The BRST Hamiltonian still is that of equation (\ref{eq:H K}), although the BRST charge that determines the contribution of the gauge-fixing fermion is that of the non-minimal sector (\ref{eq:omega nm}). 

\subsection{Gauge Fixing}

The time evolution equations (\ref{eq:dt BRST}) contain no arbitrary functions, and thus define a unique trajectory in phase space for appropriately specified initial conditions. In that sense, we can think of the action (\ref{eq:delta S BRST}) as a gauge-fixed action, with a gauge-fixing condition that is associated with the choice of $\delta K$. 
  
It is important to realize that, the choice of $\delta K$ is completely arbitrary,  up to Grassmann parity and ghost number. There is hence a huge amount of freedom to gauge fix the theory. In 
field theories in Minkowski spacetime, manifest relativistic invariance and renormalizability severely restrict the choice of the gauge fixing fermion, but in a gravitational theory in a time-dependent cosmological background the set of possible $\delta K$ is much larger, even if one insists on manifest invariance under translations and rotations. 

At this point we shall mostly focus on the simplest  gauge fixing fermions, namely, those linear in the fields of ghost number minus one,
\begin{equation}\label{eq:delta K}
	\delta K=\chi^a\, \delta\mathcal{P}_a
	+i \sigma^a \, \delta C_a.
\end{equation}
The coefficients $\chi^a$ and $\sigma^a$ are assumed to be arbitrary functions of the bosonic fields of the theory, that is, the original canonical pairs, the Lagrange multipliers and their conjugate momenta in the non-minimal sector. With this choice, by repeated use of the identities (\ref{eq:P symmetry}) and (\ref{eq:P Leibnitz}), we arrive at
\begin{multline}\label{eq:BRST correction}
\Delta\delta H\equiv\{\delta K, \delta \Omega\}=
	-\chi^a \delta G_a-\sigma^a \delta b_a
	\\
	+\delta_\eta \chi^a \delta\mathcal{P}_a
	+\chi^a \delta\eta^b C_{ba}{}^c \delta\mathcal{P}_c
	-i\delta_\rho \chi^b\delta\mathcal{P}_b
	+i\delta_\eta \sigma^a \delta C_a
	+\delta_\rho \sigma^a\delta C_a
	\\
	-\frac{1}{2}\{C_{cb}{}^a,\chi^d\}\delta\eta^b \delta \eta^c \delta\mathcal{P}_a\delta\mathcal{P}_d
	-\frac{i}{2}\delta C_a \{\sigma^a,C_{cb}{}^d\} \delta\eta^b\delta\eta^c\delta\mathcal{P}_d
	\\
	+\{\chi^a\delta\mathcal{P}_a,\delta\Omega_{p\geq 2}\}
	-i\{\sigma^a,\delta\Omega_{p\geq 2}\}\delta C_a.
\end{multline}
where $\delta_\eta \chi$ and $\delta_\rho \sigma$ respectively are the changes of $\chi$ and $\sigma$ under gauge transformations with parameters $\delta\eta^a$ and $\delta\rho^a$,  ${\delta_\eta \chi \equiv\delta\eta^a \{\delta G_a,\chi\}}$ and ${\delta_\rho \sigma \equiv\delta\rho^a \{\delta b_a,\sigma\}}$.  

It is illustrative to check how this formalism  reproduces the well-known Faddeev-Popov action in the case of canonical gauge conditions.  Setting $\chi^a=-\delta\lambda^a$ and $\sigma^a=-f^a(\delta p_i, \delta q^i)/\varepsilon$ in equation (\ref{eq:BRST correction}), and changing variables  $ \delta b_a\to\varepsilon \delta b_a$, $\delta C_a\to\varepsilon \delta C_a$ in the  gauge-fixed action (\ref{eq:delta S BRST non-min}) returns, in the limit $\varepsilon\to 0$,
\begin{multline}\label{eq:Faddeev-Popov-action}
	\delta S_\mathrm{FP}=\int dt\Big[\delta\dot{q}^i \delta p_i+\delta\dot{\eta}^a\delta\mathcal{P}_a
	-\delta H_D-\delta\lambda^a\delta G_a
	-\delta b_a f^a
	+i\delta_\eta f^a \delta C_a\\
	+i\delta\rho^a \delta\mathcal{P}_a
	+\delta\lambda^a \delta\eta^b C_{ba}{}^c \delta\mathcal{P}_c
	-\frac{i}{2}\delta C_a\{f^a,C_{cb}{}^d\}\delta\eta^b\delta\eta^c\delta\mathcal{P}_d
	+\delta\lambda^a\{\delta\mathcal{P}_a,\delta\Omega_{p\geq 2}\}+i\{f^a,\delta\Omega_{p\geq 2}\}\delta C_a
	\Big].
\end{multline}
Because the conjugate momenta $\delta b_a$  appear linearly in this action, integration over these variables produces a delta function $\delta(f^a)$ that enforces the canonical gauge-fixing conditions $f^a=0$. Hence, the first line of equation (\ref{eq:Faddeev-Popov-action}) is nothing but the gauge-fixed Faddeev-Popov  action, which includes the original Hamiltonian, the gauge-fixing term  and the corresponding ghost contribution $i\delta\eta^b\{\delta G_b,f^a\}\delta C_a$. Similarly, integration over the (imaginary) variable $\delta\rho^a$ enforces the conditions $\delta\mathcal{P}_a=0$, which eliminates all the remaining terms on the second line of the equation, since they are all of antighost number $p\geq 1$.  In that sense, the imposition of canonical gauge conditions is just a special case of gauge fixing in the BRST formalism.  Note, however, that because we are dealing with an open algebra, equation  (\ref{eq:BRST  correction}) implies that  the gauge-fixed Hamiltonian  will generically contain  cubic and higher order terms in the ghosts that do not appear in the Faddeev-Popov formalism in generalized gauges. 

Equation (\ref{eq:BRST correction}) also enables us to finally determine a BRST-invariant extension of the first-class Hamiltonian $\delta H_D$. Under the assumption that the BRST charge has the structure (\ref{eq:omega structure}), it is easy to check that $\delta H_\mathrm{BRS}\equiv-\dot{\bar q}^i \delta p_i+\dot{\bar p}_i \delta q^i$ satisfies equation (\ref{eq:BRST extension}). In the background-independent formulation of the theory, this is just the statement that an identically vanishing Hamiltonian is BRST-invariant.  Adding to the latter the BRST-exact expression obtained by setting $\chi^a=-\bar{\lambda}^a$ and $\sigma^a\equiv 0$ in equation (\ref{eq:BRST correction}) we arrive at
\begin{equation}\label{eq:BRST Hamiltonian}
	\delta H_\mathrm{BRS}\equiv 
	-\dot{\bar q}^i \delta p_i+\dot{\bar p}_i \delta q^i -\bar\lambda^a\{\delta\mathcal{P}_a,\delta\Omega\}=
	\delta H_D
	-\bar{\lambda}^b \delta\eta^c C_{cb}{}^a \delta\mathcal{P}_a
	-\bar\lambda^a\{\delta\mathcal{P}_a,\delta\Omega_{p\geq 2}\},
\end{equation}
which carries ghost number zero and satisfies the two properties (\ref{eq:BRST extension}).
The Hamiltonian (\ref{eq:BRST Hamiltonian}) shall be the starting point of our perturbative calculations. Similarly, we can construct a BRST-invariant extension of the identity operator from any gauge-fixing fermion $\delta J$ by the prescription
\begin{equation}
	\mathbbm{1}_{J}=
	\exp\left(i\{\delta J,\delta\Omega\}\right).
\end{equation}
This extension shall prove to be useful in defining BRST-invariant extensions of gauge-invariant operators.

\section{Quantization}
\label{sec:BRST Quantization}

In Dirac's  approach to the quantization of gauge theories in the Schr\"odinger representation, the Hilbert space consists of wave functionals of the configuration variables $\delta q^i$. The constraints are imposed on the physical states of the theory, $\delta\hat{G}_a |\psi_\mathrm{phys}\rangle=0$, and the time evolution operator is taken to be
\begin{equation}\label{eq:U Dirac}
\hat{\mathcal{U}}_D(t;t_0)=\mathcal{T}\exp\left[-i\int_{t_0}^{t} \delta H_D(\delta \hat q^i(t_0),\delta \hat p_i(t_0),\tilde{t}\,) \, d\tilde{t}\right],
\end{equation}
where $\mathcal{T}$ is the time-ordering operator, and we have assumed that the Hamiltonian is of the form (\ref{eq:H HD}). We shall not pursue  Dirac quantization in the Fock representation here, which proceeds by enforcing only half of the constraints on the physical states. 

BRST quantization in the Schr\"odinger representation follows basically the same steps as  Dirac quantization, basically by replacing gauge-invariance by BRST-invariance:

\begin{enumerate}
\item The real even variables in  phase space are replaced by bosonic hermitian field operators, and the real odd variables are replaced by real fermionic operators. The field operators $\delta\hat{q}^i$, $\delta\hat{p}_i$,  $\delta\hat{\eta}^a$  and $\delta \hat{C}_a$ are hermitian, while the fermionic fields $\delta\hat{\mathcal{P}}_a$ and $\delta\hat\rho^a$ in the non-minimal sector are anti-hermitian.

\item The graded Poisson brackets of the canonical variables are replaced by the graded commutators of the corresponding operators,
\begin{equation}\label{eq:graded commutations}
[\hat{q}^i,\hat{p}_j]\equiv i \delta^i{}_j,
\quad
[\delta\hat{\eta}^a,\delta\hat{\mathcal{P}}_b]\equiv -i \delta^a{}_b.
\end{equation} 
\item  Physical states $|\Psi\rangle$ belong to the cohomology of $\delta \hat{\Omega}$ at  ghost number zero,
\begin{equation}
	|\Psi\rangle\in \left(\frac{\mathrm{Ker}\, \delta\hat\Omega}{\mathrm{Im}\, \delta\hat\Omega}\right)_0.
\end{equation}

\item The gauge-fixed Hamiltonian $\delta H_K$ generates time evolution,
\begin{equation}
i\frac{d\delta \hat F}{dt}=[\delta\hat F, \delta\hat{H}_K]+i\frac{\partial \delta \hat F}{\partial t}.
\end{equation}
\end{enumerate}
Hence, the Heisenberg representation fields at time $t$ can be recovered from the time-evolution operator as usual,
\begin{align}
	\delta \hat F(t)&=\hat{\mathcal{U}}_K^\dag(t,t_0) \delta \hat F(t_0) \hat {\mathcal{U}}_K(t,t_0),
	\\ 
	\hat{\mathcal{U}}_K(t,t_0)&\equiv \mathcal{T}\exp\left(-i\int_{t_0}^{t} \, \delta\hat{H}_K\, d\tilde{t} \right). \label{eq:BRST U}
\end{align}
 In this formulation time evolution is  manifestly unitary. In the Schr\"odinger picture, it is also manifestly insensitive to the gauge-fixing fermion $\delta \hat{K}$, because the latter only changes the  physical state $|\Psi\rangle$ by vectors in the image of the BRST charge, which do not affect the cohomology of $\delta\hat{\Omega}$. In the following we remove the hats on top of the operators for simplicity.

\subsection{Expectation Values}

\subsubsection{BRST Quantization} In cosmological perturbation theory, we are  interested in the expectation values of observables in appropriately chosen quantum states. A convenient basis  is furnished by the BRST-invariant states of ghost number zero
\begin{equation}\label{eq:BRST inv}
	|\delta q^i_\mathrm{BRS}\rangle=| \delta q^i, \delta \eta^a=0, \delta b_a=0, \delta C_a=0\rangle,
\end{equation}
where the labels indicate the eigenvalues of the corresponding operators.  An important theorem in BRST theory\footnote{See Section 14.5.5 in \cite{Henneaux:1992ig}.} states that if $\delta \mathcal{O}_\mathrm{BRS}$ is a BRST-invariant extension of a given observable $\delta\mathcal{O}$, its matrix elements between BRST-invariant states of the form (\ref{eq:BRST inv}) return the projected kernel of $\delta \mathcal{O}$, up to an irrelevant overall sign, 
\begin{equation}\label{eq:projected kernel}
	\langle \delta q^i_\mathrm{BRS}| \delta \mathcal{O}_\mathrm{BRS}| \delta \tilde q_\mathrm{BRS}^j\rangle = \delta \mathcal{O}_P(\delta q^i, \delta\tilde q^j). 
\end{equation}
The projected kernel $\delta \mathcal{O}_P$ on the right hand side is a regularized version of the kernel of $\delta\mathcal{O}$  projected onto the space of gauge-invariant states.  As a consequence of this projection,  $\delta\mathcal{O}_P$ satisfies the constraints, in the sense that 
\begin{equation}
	\delta G_a(\delta q^i, -i\partial/\partial \delta q^i)\delta\mathcal{O}_P(\delta q^i, \delta\tilde q^j)=
	\delta G_a(\delta\tilde  q^i, -i\partial/\partial \delta\tilde q^i)
	\delta\mathcal{O}^*_P(\delta q^i, \delta\tilde q^j)=0.
\end{equation}
In general, in order for the matrix element (\ref{eq:projected kernel}) to be well-defined, it is necessary that the BRST-invariant extension $\delta\mathcal{O}_\mathrm{BRS}$ mix the minimal and non-minimal sectors.   With these definitions, the expectation value of a gauge-invariant operator $\delta\mathcal{O}(t)$ becomes
\begin{subequations}\label{eq:diagonal vev all}
\begin{equation}\label{eq:diagonal vev}
	\langle \Psi_\mathrm{in} | \delta \mathcal{O}(t) | \Psi_\mathrm{in}\rangle
	=
	\int d\delta q \, d\delta \tilde q\,
	\psi^*_\mathrm{in}(\delta q) \mu(\delta q)\,
	\delta\mathcal{O}_P(t)(\delta q,\delta \tilde{q})
	\mu(\delta\tilde q)  \,
	\psi_\mathrm{in}(\delta \tilde q),
\end{equation}
where the projected kernel of the operator equals, using equation (\ref{eq:projected kernel}),
\begin{equation}\label{eq:vev extension}
	\delta \mathcal{O}_P(t)(\delta q, \delta \tilde{q})=
	\langle \delta q_\mathrm{BRS}| \, 
	\big[\mathcal{U}_D^\dag(t,t_0) 
	\delta \mathcal{O}(t_0) 
	\mathcal{U}_D(t,t_0)\big]_\mathrm{BRS}\, 	
	 | \delta \tilde q_\mathrm{BRS} \rangle.
\end{equation}
In equation (\ref{eq:diagonal vev}), $\mu$ is a  regularization functional needed to render  the integrals finite\footnote{\textit{ibid.}, Section 13.3.4.}. In simple cases in which the constraints are linear in the conjugate momenta  it is determined by  a set of conditions $f^a(\delta q,\delta \pi)=0$ through the relation 
\begin{equation}\label{eq:mu}
	\mu=\det\{f^a,\delta G_b\}\delta(f^c).
\end{equation}
In the absence of this regularization factor, the integral over the gauge-invariant directions would diverge, since  the projected kernel satisfies the constraints. We illustrate this divergence with a concrete example in Section \ref{sec:Quantization: Vectors}.  Although the constraints in general relativity are not linear in the momenta, we shall only need  the form of $\mu$ in the limit in which interactions are turned off, and the form (\ref{eq:mu}) does apply.  
 
 The BRST extension of $\mathcal{U}_D^\dag(t,t_0) \delta \mathcal{O}(t_0) \mathcal{U}_D(t,t_0)$ in equation (\ref{eq:vev extension}) is pretty much arbitrary, up to the requirement that it mix the minimal and non-minimal sectors. A relatively natural choice would be 
$
	\mathcal{U}_K^\dag(t,t_0) 
	\delta\mathcal{O}(t_0)_\mathrm{BRS} \,
	\mathcal{U}_K(t,t_0),
$ 
for a $\delta K$ that mixes the minimal and non-minimal sectors. This is the choice typically made in approaches in which one is interested in calculating in-out matrix elements of the time evolution operator. 
But since we are interested in in-in matrix elements,   in some instances this prescription fails    because the terms that mix  sectors in $\mathcal{U}^\dag_K$ and $\mathcal{U}_K$ cancel each other. It  is thus convenient to work instead with the slightly more general expression
\begin{equation}\label{eq:observable extension}
[\mathcal{U}_D^\dag(t,t_0) 
	\delta \mathcal{O}(t_0) 
	\mathcal{U}_D(t,t_0)]_\mathrm{BRS}=
	\mathbbm{1}_J(t_0)\,
	\mathcal{U}_K^\dag(t,t_0) 
	\delta\mathcal{O}_\mathrm{BRS}(t_0) \,
	\mathcal{U}_K(t,t_0)\,
	\mathbbm{1}_J(t_0),
\end{equation}
\end{subequations}
for appropriately chosen and unrelated gauge-fixing fermions $\delta J$ and $\delta K$. The formalism guarantees that the projected kernel is insensitive to the particular choice of BRST extension. Although  it is not necessary to include two copies of $\mathbbm{1}_J$ in the extension (\ref{eq:observable extension}), their inclusion renders it slightly more symmetric.  
  
\subsubsection{Dirac Quantization}

Within the BRST formalism it is also possible to recover the  matrix elements of the Dirac time evolution operator (\ref{eq:U Dirac}) from those of the BRST-invariant operator in equation (\ref{eq:BRST U}). In those cases, it is not necessary to choose a gauge-fixing fermion that mixes the minimal and non-minimal sectors, and one can indeed set $\delta K=0$. With a vanishing gauge-fixing fermion, the Hamiltonian  of the theory reduces to the first-class Hamiltonian $\delta H_D$, plus the ghost terms needed to construct its BRST extension. Since the  first-class Hamiltonian $\delta H_D$  is that of the original theory with Lagrange multipliers $\delta\lambda^a$ set to zero, the bosonic sector of the action  is the one in equation (\ref{eq:Hamiltonian S}) for a metric of the form 
\begin{equation}\label{eq:synchronous coords}
	ds^2=-\bar{N}^2 dt^2+(\bar{h}_{ij}+\delta h_{ij})dx^i dx^j.
\end{equation}
Up to the irrelevant factor $\bar{N}^2=a^2$, this is just the metric in synchronous coordinates, or 
synchronous gauge. 

But if one is interested in the  unprojected kernel of the time evolution operator in the Dirac method anyway, one can dispense of ghosts and BRST-invariance altogether, and simply rely on the original representation of the operator in equation (\ref{eq:U Dirac}).  Quantization of cosmological perturbations following Dirac's method thus amounts to quantization of the theory with a metric of the form (\ref{eq:synchronous coords}), with no ghosts in the action.  For that reason, we shall use ``Dirac quantization" and ``synchronous gauge quantization" as synonyms, in spite of our previous remarks concerning synchronous gauge in the BRST formalism.  In particular, in synchronous  gauge the expectation value of a gauge-invariant operator reads
\begin{subequations}\label{eq:non-diagonal vev all}
\begin{equation}\label{eq:non-diagonal vev}
	\langle \Psi_\mathrm{in} | \delta \mathcal{O}(t) |  \Psi_\mathrm{in}\rangle
	=
	\int d\delta q \, d\delta \tilde q\,
	\psi^*_\mathrm{in}(\delta q) \mu(\delta q) 
	\delta \mathcal{O}(t)(\delta q,\delta \tilde{q})
	\psi_\mathrm{in}(\delta \tilde q),
\end{equation}
where the kernel of the operator $\delta \mathcal{O}(t)$ equals
\begin{equation}\label{eq:O Heisenberg}
\delta \mathcal{O}(t)(\delta q,\delta \tilde{q})=
	\langle \delta q | 
	\mathcal{U}^\dag_D(t,t_0) 
	\delta \mathcal{O}(t_0)\,
	\mathcal{U}_D(t,t_0)	
	 | \delta \tilde q \rangle,
\end{equation}
\end{subequations}
 and the time evolution operator is that of equation (\ref{eq:U Dirac}). Note that it is only necessary to insert a single regularization factor $\mu$ in the integrand of (\ref{eq:non-diagonal vev}), since it contains the non-projected kernel of the operator $\delta\mathcal{O}$.  The factor of $\mu$ then regularizes the scalar product between the gauge-invariant states $|\Psi_\mathrm{in}\rangle$ and $\delta\mathcal{O}(t)|\Psi_\mathrm{in}\rangle$. Clearly, one of the main advantages of Dirac (or synchronous gauge) quantization is the absence of ghosts. 

\section{Free Theory}

Our discussion of gauge-invariance and BRST symmetry so far has been mostly exact. Even though we have expanded  the fields around a non-trivial cosmological background, the expressions that we have derived apply to all orders in the perturbations. In practice, however, such calculations do not occur. Instead, one typically expands the action to a certain order in the perturbations, and carries out calculations only to that order. In that case, all the expressions we have previously derived hold only to the appropriate order in the perturbation expansion. As we shall see, such perturbative calculations  obscure the  symmetries and simplicity of the underlying theory even further. 

We shall first  illustrate such a  calculation in the simplest case, that of linear perturbation theory. This case is  relevant because it provides the foundation upon which perturbative calculations are built, and because it suffices  to  calculate the primary observables of inflationary theory, namely, primordial power spectra. The linear analysis can be easily generalized to  arbitrary higher orders, at least formally, although specific calculations become increasingly cumbersome as the perturbation order increases.

\subsection{Einstein-Hilbert Action}

We begin by expanding the Einstein-Hilbert action in Hamiltonian form (\ref{eq:S pert I})  to quadratic order in the perturbations. The action therefore becomes 
\begin{equation}\label{eq:delta S Hamiltonian}
	\delta S=\int dt\left[\delta p_i \delta\dot{q}^i-\delta^{(2)}\!H\right],
\end{equation}
where $\delta^{(2)}H$ is the Hamiltonian expanded to second order,
\begin{equation}
\delta^{(2)}\!H=-\dot{\bar{q}}^i \delta p_i+ \dot{\bar{p}}_i \delta q^i+\bar{\lambda}^a\delta^{(2)} G_a+\delta\lambda^a\delta^{(1)}G_a,
\end{equation}
and the $\delta^{(n)}G_a$ are the constraints expanded to $n$-th order in the perturbations. Note that the Hamiltonian somewhat simplifies because the background satisfies the classical equations of motion, so the linear terms in the perturbations cancel.  Hence, if $\delta G^{(k)}$ is the term of $k$-th order in $\delta G$,
we can also write
\begin{equation}
	\delta^{(2)}\!H=\bar\lambda^a \delta G_a^{(2)}+\delta\lambda^a \delta G_a^{(1)}\equiv \delta^{(2)}\!H_D+\delta\lambda^a \delta G_a^{(1)},
\end{equation}

In order to find $\delta G_a^{(2)}$ and $\delta G_a^{(1)}$, we simply expand  expressions (\ref{eq:secondary}) to the desired order.  We begin with the terms linear in the perturbations, 
\begin{subequations}\label{eq:linear constraints}
\begin{align}
	\delta G^{(1)}_N &=
	2a \mathcal{H}\delta\pi
	+\frac{\dot{\bar\varphi}}{a}\delta\pi_\varphi
	+\frac{M^2}{2a}\left(\nabla^2 \delta h-\partial_i\partial_j \delta h^{ij}\right)
	+\frac{M^2}{a}\dot{\mathcal{H}}\delta h+a^3 \bar{V}_{,\varphi}\delta\varphi,\\
	\delta G^{(1)}_i&=-\left[2a^2\partial_j\delta\pi_i{}^j-M^2 \mathcal{H}(2\partial_j \delta h_i{}^j-\partial_i \delta h)\right]
	+a^2\dot{\bar\varphi}\, \partial_i\delta\varphi,
\end{align}
\end{subequations}
in which indices are raised and lowered  with a Kronecker delta. To derive these expressions, we have used the background equations of motion (\ref{eq:background motion}), and that the relevant background quantities  satisfy (\ref{eq:background}).

Because in our background the shift vector vanishes ($\bar{\lambda}^i\equiv 0$), in order to obtain the quadratic Hamiltonian we only need to calculate $\delta G_N$ to second order. After a somewhat long but straight-forward calculation we arrive at 
\begin{multline}\label{eq:H0 pert}
	\delta H_D^{(2)}=\int d^3 x\bigg[\frac{2a^2}{M^2}\delta\pi^{ij}\delta\pi_{ij}-\frac{a^2}{M^2}\delta\pi^2+\frac{1}{2a^2}\delta\pi_\varphi^2
	-2\mathcal{H}\delta\pi^{ij}\delta h_{ij}+\mathcal{H}\delta\pi\delta h-\frac{\dot{\bar\varphi}}{2a^2}\delta\pi_\varphi \delta h\\
	 -\frac{M^2}{8a^2}\delta h^{ij}\nabla^2 \delta h_{ij}+\frac{M^2}{8a^2}\delta h\nabla^2 \delta h+\frac{M^2}{4a^2}\delta h^{ij} \partial_j\partial_k \delta h_i^k-\frac{M^2}{4a^2}\delta h \partial_j \partial_k \delta h^{jk}
	\\
	+\frac{a^2}{2}\partial_i\delta\varphi \partial^i\delta\varphi
	+\frac{M^2\left(\mathcal{H}^2-\dot{\mathcal{H}}/2\right)}{a^2}\delta h_{ij}\delta h^{ij} +\frac{a^4}{2}\bar{V}_{,\varphi\varphi}\delta\varphi^2+\frac{a^2 \bar{V}_{,\varphi}}{2}\delta\varphi\delta h
	\bigg].
\end{multline}

\subsection{Constraints}

Variation of the action (\ref{eq:delta S Hamiltonian}) with respect to $\delta\lambda^a$  results in the linear constraints $\delta G_a^{(1)}=0$, or, in standard notation,
\begin{equation}
	\delta G_N^{(1)}=\delta G_i^{(1)}=0.
\end{equation}
The reader can readily check that the linear constraints (\ref{eq:linear constraints}) generate linear diffeomorphism when acting on the configuration variables. In particular, $\delta G_N^{(1)}$ generates linear diffeomorphisms along the unit normal to the constant time hypersurfaces $n^\mu=(a^{-1},\vec{0})$. But  as opposed to that of the full theory,  the algebra of the linearized constraints (\ref{eq:linear constraints}) is Abelian. This  follows for instance by expanding equation (\ref{eq:structure}) in powers of the perturbations, and noting that the background satisfies the zeroth order constraints. 

Because the full action for the perturbations (\ref{eq:S pert I})  is invariant under the transformations (\ref{eq:Delta delta}), the quadratic action (\ref{eq:delta S Hamiltonian}) is invariant under a truncation of those transformations to linear order,
\begin{subequations}\label{eq:Delta delta linear}
\begin{align}
	\Delta \delta q^i&=\{\delta q^i,\xi^a \delta G^{(1)}_a\}, \\
	\Delta \delta p_i&=\{\delta p_i,\xi^a \delta G^{(1)}_a\},\\
	\Delta \delta \lambda^a&=\dot{\xi}^a
	+\xi^b\bar{\lambda}^c \bar{C}_{bc}{}^a.
	\label{eq:Delta delta lambda linear}
\end{align}
\end{subequations}
Readers familiar with  gauge transformations in the Hamiltonian formalism will recognize in equation (\ref{eq:Delta delta lambda linear})  the transformation properties of the Lagrange multipliers under a  set of Abelian first class constraints with a weakly gauge-invariant Hamiltonian.  

It is in fact illustrative to see how  the linear constraints 
(\ref{eq:linear constraints}) are preserved by the time evolution. On the one hand, from the equations of motion in the linearized theory, their time derivatives are
\begin{equation}
	\frac{d\delta G_a^{(1)}}{dt}\equiv \{\delta G_a^{(1)},\delta^{(2)}\!H\}+\frac{\partial\delta G_a^{(1)}}{\partial t}.
\end{equation}
On the other hand, in the theory to all order in the perturbations, the time derivatives of the full constraints obey, from equations (\ref{eq:structure}) and (\ref{eq:dt delta C}),
\begin{equation}\label{eq:full evolution}
	\frac{dG_a}{dt}\equiv \{\delta G_a,\delta H\}+\frac{\partial\delta G_a}{\partial t}=(\bar\lambda^b+\delta\lambda^b) C_{ab}{}^c \delta G_c.
\end{equation}
Expanding the equation on the right hand side of (\ref{eq:full evolution}) to first order in the perturbations, and bearing in mind that $\delta H^{(1)}$ vanishes, we immediately get that 
\begin{equation}\label{eq:dGa 1 a}
	 \{\delta G_a^{(1)},\delta^{(2)}\!H\}+\frac{\partial\delta G_a^{(1)}}{\partial t} =\bar\lambda^b \bar{C}_{ab}{}^c  \delta G_c^{(1)}.
\end{equation}
Therefore, the time derivatives of the linear constraints  vanish weakly, and the Hamiltonian itself is not  invariant under diffeomorphism, even on-shell. In the full theory, the linearity of the Hamiltonian in the first class constraints is essentially responsible for  the preservation of the constraints under the time evolution. In the linear theory, this simple structure is hidden  and distorted by the expansion  to second order in the perturbations around a time-dependent background. Note that the coefficient $\bar\lambda^b \bar{C}_{ab}{}^c$ multiplying the constraints on the right hand side of (\ref{eq:dGa 1 a}) is precisely the coefficient that appears in the transformation law (\ref{eq:Delta delta lambda linear}). This is what one expects in a theory in which the Hamiltonian is weakly gauge-invariant.

\subsection{Classical BRST Symmetry}

The BRST charge of the linearized theory (in the non-minimal sector) is that of the full theory, equation (\ref{eq:omega nm}), expanded to second order in the  perturbations,
\begin{equation}\label{eq:delta omega linear}
	\delta\Omega ^{(2)}\equiv \delta\eta^a \delta G_a^{(1)}-i\delta\rho^a\delta b_a.
\end{equation}
Again, this is consistent with our observation that the constraints in the linear theory are Abelian. In order to write down the BRST-invariant action in the free theory, we need to find the BRST Hamiltonian. Expanding equation (\ref{eq:BRST Hamiltonian}) to second order, we readily arrive at 
\begin{equation}\label{eq:delta 2 H BRST}
	\delta^{(2)}\!H_\mathrm{BRS}\equiv \delta^{(2)}\!H_D-\bar{\lambda}^a \delta\eta^b \bar{C}_{ba}{}^c\delta\mathcal{P}_c,
\end{equation}
which  indeed is BRST-invariant, in the sense that it obeys
\begin{equation}\label{eq:Omega 1 conserved}
 	\{\delta\Omega^{(2)},\delta^{(2)}H_\mathrm{BRS}\}+\frac{\partial\delta\Omega^{(2)}}{\partial t}=0,
\end{equation}
as the reader can verify using equation (\ref{eq:dGa 1 a}).  The BRST-invariant action of the linearized theory in the non-minimal sector still has the form (\ref{eq:delta S BRST non-min}), but the gauge-fixed Hamiltonian is that obtained from equation (\ref{eq:delta 2 H BRST}) and the usual relation
\begin{equation}
\delta^{(2)}\!H_K=\delta^{(2)}\!H_\mathrm{BRS}+\{\delta K,\delta\Omega^{(2)}\},
\end{equation}
 where the gauge-fixing fermion $\delta K$ is an arbitrary Grassmann-odd function of the canonical variables of ghost number minus one. The gauge-fixed Hamiltonian still satisfies  equation (\ref{eq:Omega 1 conserved}), which  guarantees that, up to total derivatives, the quadratic action  is invariant under the linear BRST transformations generated by $\delta\Omega^{(2)}$.   To preserve the quadratic structure of the Hamiltonian, $\delta K$ needs to be quadratic in the perturbations too. 

\subsubsection*{Comohology}

In a theory with a gauge symmetry the algebra of observables consists of gauge invariant functions of the canonical variables. As we mentioned earlier, this algebra agrees with the set of BRST-invariant functions of ghost number zero. To see how this works in practice, let us determine the cohomology of the linearized BRST charge (\ref{eq:delta omega linear}).  

We begin by calculating the action of $\delta\Omega^{(2)}$ on the different fields in the non-minimal sector of the theory,
\begin{align}
	s\, \delta\lambda^a&=-i\delta\rho^a, 
	&s\,\delta\rho^a&=0,
	\\
	s\,\delta C_a&=i\delta b_a,  &s \, \delta b_a&=0.
\end{align}
Since all the fields that are closed ($s\delta F=0$) are exact ($\delta F=s\delta G$), the cohomology  in the non-minimal sector is trivial. Hence, observables can be assumed not to depend on the variables of the non-minimal sector. In particular, they  do not depend on the Lagrange multipliers  $\delta\lambda^a\equiv\{\delta \lambda^N,\delta\lambda^i\}$. 

In the minimal sector, the action of the BRST charge amounts to a gauge transformation with gauge parameters $\delta\eta^a$.  To study the cohomology, it proves to be useful to move to a different field basis. First, we carry out  a canonical transformation to a set of fields that transform irreducibly, as described in Appendix \ref{sec:Irreducible Representations}. The latter are classified according to  their transformation properties under rotations as scalars, vectors and tensors.  In linear perturbation theory the three sectors decouple from each other. In each sector, then, we perform another canonical transformation to a basis of fields with particularly simple properties under gauge transformations.  

\paragraph{Tensor Sector} In the tensor sector this basis consists of the conjugate pairs
\begin{equation}
	\zeta_{\pm2} \equiv \frac{ \delta h_{\pm2}}{a^2}, \quad \Pi^{\pm2}\equiv a^2\delta\pi^{\pm2}.
\end{equation}
Because they are invariant under the linearized diffeomorphisms, and there is no helicity two field of ghost number minus one, the helicity two fields are exact  and cannot be written as a BRST transformation of any other fields. They span the observables of the tensor sector of the linear theory, and the BRST charge in this sector identically vanishes. The Hamiltonian (\ref{eq:H0 pert})  in the tensor sector reads
\begin{equation}
	\delta H^{(2)}_{D,t}=
	\frac{1}{2}\sum_{\sigma=\pm 2}\int d^3 p 
	\left[
	\frac{\Pi^\sigma (\vec{p})\Pi^\sigma (-\vec{p})}{ a^2 M^2}
	-8\mathcal{H}\,\Pi^\sigma(\vec{p})\zeta_\sigma(\vec{p})
	+a^2 M^2\left(\vec{p}\,^2
	+8\mathcal{H}^2-4\dot{\mathcal{H}}\right)\zeta_\sigma(\vec{p})\zeta_\sigma(-\vec{p})\right].
\end{equation}
There is no contribution from the tensor sector to the BRST-charge,
$
	\delta\Omega^{(2)}_t=0.
$

\paragraph{Vector Sector} In the vector sector the new  field basis is
\begin{equation}
	x^{\pm 1}\equiv -\frac{\delta h_{\pm1}}{a^2 p}(\vec{p}), 
	\quad
	G_{\pm1}\equiv 
	-p\left[a^2\delta\pi^{\pm1}(\vec{p})-2M^2 \mathcal{H} \delta h_{\pm1}(-\vec{p})\right].
\end{equation}
Among other properties, this basis is useful because the constraints in the vector sector simply read $G_{\pm1}=0$.  The action of a BRST transformation on each of these helicity one fields is
\begin{subequations} 
\begin{align}
s \, G_{\pm1} &=0,
\quad
&s\,\delta\mathcal{P}_{\pm1}&= -G_{\pm1},\\
s \, \delta\eta^{\pm1}&=0, &s\, x^{\pm1}&=\delta\eta^{\pm1}. 
\end{align} 
\end{subequations}
Hence, all BRST-closed fields are BRST-exact. Clearly, the BRST cohomology in the vector sector is trivial and  there are no observables in this sector.  The Hamiltonian is
\begin{equation}\label{eq:Hv}
	\delta H^{(2)}_{D,v}=\sum_{\sigma=\pm 1} \int d^3 p\,
	\bar{A}_\sigma G_\sigma (\vec{p})G_\sigma (-\vec{p})
	\quad \bar{A}_{\pm1}=\frac{1}{a^2 M^2 p^2},
\end{equation}
and  the BRST charge in the vector sector reads
\begin{equation}\label{eq:Omegav}
	\delta\Omega^{(2)}_v=\sum_{\sigma=\pm1} \int d^3 p \,\big[\delta \eta^\sigma(\vec{p}) G_\sigma(\vec{p})-i\delta\rho^\sigma (\vec{p}) \delta b_\sigma(\vec{p})\big].
\end{equation}
Since the Poisson bracket of the first-class Hamiltonian with the BRST charge  vanishes, $\delta H^{(2)}_{D,v}$  is BRST-invariant. 

\paragraph{Scalar Sector} The bosonic scalar sector is spanned by the three canonical pairs  $(\delta h_L,  \delta\pi^L), (\delta h_T, \delta\pi^T)$ and $(\delta\varphi, \delta\pi_\varphi)$.  A new choice of conjugate pairs  that happens to be particularly convenient is
$(\zeta,\Pi), (x^L,G_L), (x^N, G_N)$, where the latter are defined by
\begin{subequations}\label{eq:new scalar variables} 
\begin{align}
	\zeta &\equiv\frac{\delta h_L}{a^2}+\frac{\delta h_T}{3a^2}-\frac{\mathcal{H}}{\dot{\bar{\varphi}}}\delta\varphi,
	\\
	\Pi &\equiv
	\frac{6 M^2 \mathcal{H}^2-2M^2p^2}{\mathcal{H}}\delta h_L
	-\frac{2 M^2p^2}{3\mathcal{H}}\delta h_T
	-3a^2\dot{\bar\varphi}\,\delta\varphi
	+a^2\delta\pi^L
	\\
	 x^L&\equiv-\frac{\delta h_T}{p a^2}, 
	 \\
	  G_L &\equiv 2\mathcal{H}M^2 p\, \delta h_L+\frac{8}{3} \mathcal{H} M^2 p\, \delta h_T
	  -a^2 \dot{\bar\varphi}\, p \, \delta\varphi
	  +\frac{a^2}{3} p\, \delta\pi^L-a^2 p\,  \delta \pi^T,
	  \\
	 x^N&\equiv \frac{a \,\delta\varphi}{\dot{\bar\varphi}}
	 ,\\
	 G_N &\equiv -\frac{2M^2}{a}\left(p^2-3\dot{\mathcal{H}}\right)\delta h_L-\frac{2M^2 p^2}{3a}\delta h_T+a^3 \bar{V}_{,\varphi} \delta\varphi+a\mathcal{H}\,\delta \pi^L+\frac{\dot{\bar\varphi}}{a}\delta\pi_\varphi.
\end{align}
\end{subequations}
We quote the inverse relations that express the old variables in terms of the new variables in equations (\ref{eq:inverse relations}).

The new variables $G_L$ and $G_N$ are nothing but the scalar components of the original linearized constraints (\ref{eq:linear constraints}), which now simply read $G_L=G_N=0$. The variable $\zeta$ is the standard curvature perturbation in comoving slices, and  setting $x^L=x^N=0$ amounts to working in comoving gauge. The variables $\zeta$, $\Pi$, as well as the constraints $G_L$ and $G_N$ are gauge-invariant. The configuration fields canonically conjugate to the latter, $x^L$ and $x^N$,  then span the two independent gauge variant directions in field space.  The action of a BRST transformation on these fields is
\begin{subequations}\label{eq:BRST scalar delta}
\begin{align}
	s\,\zeta &=0,\\
	s\,\Pi&=0, \\
	s\, G_N&=0, &s\, \mathcal{P}_N&=- G_N\\
	s\, G_L&=0, &s\, \mathcal{P}_L&=- G_L,\\
	s\,  \eta^L&=0, &s\,  x^L&=\delta\eta^L,\\
	s\,  \eta^N&=0, & s\, x^N&= \delta\eta^N.
\end{align}
\end{subequations}
Hence, the only BRST-closed fields in our set that are not BRST-exact are $\zeta$ and $\Pi$. The latter are thus the only observables in the scalar sector of the linear theory. Note that we did not have to invoke the zero ghost number condition to identify the space of observables. To conclude, let us write down the contribution of the scalar sector to the BRST charge in the new field basis,
\begin{equation}\label{eq:Omega s}
\delta\Omega^{(2)}_s=\sum_{\sigma=L,N}\int d^3 p  \big[\delta \eta^\sigma(\vec{p})  G_\sigma(\vec{p})
-i \delta\rho^\sigma(\vec{p})\delta b_\sigma (\vec{p})\big], 
\end{equation}
from which we could also have easily derived equations (\ref{eq:BRST scalar delta}). In terms of the new canonical fields, the  Hamiltonian (\ref{eq:H0 pert}) in the scalar sector   becomes
\begin{multline}\label{eq:Hs}
\delta H_{D,s}^{(2)}= \int d^3p \, \bigg[
	\frac{\mathcal{H}^2}{2a^2  \dot{\bar{\varphi}}^2} \Pi^2
	+ \frac{p^2 a^2 \dot{\bar{\varphi}}^2}{2\mathcal{H}^2} \zeta^2
	+\frac{3}{4M^2p^2 a^2}G_L^2
	+\frac{1}{2 \dot{\bar{\varphi}}^2}G_N^2
	\\
	-\frac{p}{\mathcal{H}}\zeta\,G_L
	-\Pi\left(\frac{G_L}{2M^2 a^2 p}+\frac{\mathcal{H}G_N}{a \dot{\bar{\varphi}}^2}\right)-\frac{p}{a}x^N G_L
	\bigg].
\end{multline}
In this form, the Poisson brackets of the Hamiltonian with the constraints can be read off  immediately.  Because the former do not vanish, the Dirac Hamiltonian is not BRST-invariant. Instead, from equations (\ref{eq:Omega s}) and (\ref{eq:Hs}), or directly from (\ref{eq:dGa 1 a}) and (\ref{eq:delta 2 H BRST}),  a BRST-invariant Hamiltonian is
\begin{equation}
	\delta H^{(2)}_{\mathrm{BRS},s}=\delta H_{D,s}^{(2)}-\int d^3 p \, \frac{p}{a}\delta\eta^N \delta\mathcal{P}_L.
\end{equation}
For simplicity, we have dropped the momentum labels of the different fields in the previous integrals. The latter can be reintroduced by demanding that the corresponding expression be a scalar under translations.  Enforcing the constraints on the scalar sector action by setting $G_L= G_N=0$  yields the Hamiltonian  action for the gauge-invariant conjugate variables $\zeta$ and $\Pi$,
\begin{equation}\label{eq:S zeta R}
	S_\mathrm{GI}[\Pi,\zeta]=\int d^3p \, \left[
	\Pi(\vec{p})\dot{\zeta}(\vec{p})-
	\frac{\mathcal{H}^2}{2a^2  \dot{\bar{\varphi}}^2} \Pi(\vec{p})\Pi(-\vec{p})
	- \frac{a^2 p^2 \dot{\bar{\varphi}}^2}{2\mathcal{H}^2} \zeta(\vec{p})\zeta(-\vec{p})\right].
\end{equation}
This is the action one would use in the standard gauge-invariant (or reduced phase space) approach, in which only a complete set of gauge-invariant variables is kept in the  theory. Replacing $\Pi$ by the solution of its own equation of motion yields the well-known Lagrangian action for the gauge-invariant variable $\zeta$ in the gauge-invariant formalism \cite{Garriga:1999vw}. For an alternative discussion of the reduced phase space  in linearized cosmological perturbation theory, see \cite{Langlois:1994ec}.

The transition to the new variables (\ref{eq:new scalar variables}) considerably simplifies the structure of the scalar Hamiltonian, which now is  almost diagonal. The Hamiltonian (\ref{eq:Hs})  does not depend on  $x^L$, and only depends on $x^N$ through the combination $x^NG_L$, which manifestly shows that the constraints $G_L$ and $G_N$ are preserved by the time evolution. It is in fact possible to fully diagonalize the Dirac Hamiltonian by an appropriate canonical transformation.  We discuss the diagonalization of the scalar Hamiltonian in Appendix \ref{sec:Diagonalization of the Scalar Hamiltonian}, where we show that it can be cast in the form 
\begin{equation}\label{eq:Hs diagonal}
\delta H_{D,s}^{(2)}=\int d^3p 
	\left[
	\frac{\mathcal{H}^2}{2a^2 \dot{\bar\varphi}^2} \underline{\Pi}^2
	+\frac{p^2 a^2 \dot{\bar{\varphi}}^2}{2\mathcal{H}^2} \underline{\zeta}^2
	+ \bar{A}_L\, \underline{G}_L^2
	+ \bar{A}_N\, \underline{G}_N^2
	\right],
\end{equation}
where $\bar{A}_L$ and $\bar{A}_N$ are the time-dependent coefficients quoted in equations (\ref{eq:As}).  The variable $\underline{\zeta}$ is still the comoving curvature, but only when the constraints $\underline{G}_L=\underline{G}_N=0$ are satisfied.

\subsection{Quantization: Vectors}
\label{sec:Quantization: Vectors}

The vector sector in the free theory offers a simple setting to illustrate the methods behind Dirac and BRST quantization, and how they differ from other common approaches. It will also serve as a useful warm-up for  the quantization of the scalar sector. The Hamiltonian in the vector sector is given by equation (\ref{eq:Hv}), and the BRST charge by equation (\ref{eq:Omegav}). In this sector the constraints read $G_\pm=0$.

\subsubsection{Dirac Quantization}

As we mentioned above, Dirac quantization amounts to the quantization of the theory  in synchronous gauge.  In the Dirac approach to the quantization of first class constraints, the generator of the time evolution is  the first-class Hamiltonian $\delta H_D$ in equation (\ref{eq:Hv}). The constraints are then imposed on the physical states of the theory by demanding ${G_{\pm 1}(t_0) |\Psi\rangle=0}$. If we represent these states by wave functionals in configuration space $\psi[x^{+}(\vec{p}),x^{-}(\vec{p})]$, this implies that the latter do not depend on the fields $x^{\pm1}\equiv x$. 

Observables  are functions of   gauge-invariant operators alone. The only such operators in the vector sector are the  $G_\pm$.  Since  the Hamiltonian of the theory commutes with $G_\pm$, the action of these operators on any physical state will hence vanish. But one needs to be careful when evaluating expectation values of gauge-invariant operators that do not involve powers of $G_\pm(t)$, because the latter are naively ill-defined. Consider for example the expectation of the (gauge-invariant) identity operator in the vector sector for a physical state $|\Psi\rangle$ with wave function $\psi$, a.k.a. the norm, $\int dx \,  \psi^*[x] \psi[x].$ If the state is physical, its wave function  does not depend on $x$, and the integral diverges. In order to avoid this problem, one needs to regularize the inner product  by inserting an appropriate regularization factor $\mu$, as in equation (\ref{eq:non-diagonal vev}). The latter typically involves choosing a slice through the space of gauge-variant configurations. In the present context $\mu$  can be  chosen to simply be $\delta[x]\equiv \delta[x_+]\delta[x_-]$. In that case the expectation value becomes
\begin{equation}
	 \langle \Psi | \mathbbm{1}|\Psi\rangle \equiv\int dx \,  \psi^*[x]\delta[x]\psi[x]=|\psi[0]|^2, 
\end{equation}
which equals one, as it should, if $\psi$ is properly normalized. 

\subsubsection{BRST Method}

Things look quite different in BRST quantization. We already know that in this sector the  cohomology of fields is trivial, so there are no non-trivial observables in the vector sector. Nevertheless, for illustration, let us proceed with the calculation of  expectation values outlined by equations (\ref{eq:diagonal vev all}).  One begins by choosing a convenient BRST-invariant extension of the observable at hand. For illustration we choose  the vector power spectrum, 
\begin{equation}\label{eq:vector power}
	\delta\mathcal{O}_\mathrm{BRS}(t)=
	G_+(t,\vec{p}_2)G_+(t,\vec{p}_1),
\end{equation}
which already is BRST-invariant in the free theory. We choose next a BRST extension of the time-evolution operator. Since the Hamiltonian (\ref{eq:Hv}) is already BRST-invariant, we simply set $\delta K=0$ in equation (\ref{eq:observable extension}). If we had chosen for instance $\delta K=-\delta\lambda^a \delta G_a$, the contributions from the gauge-fixing fermion  in $\mathcal{U}^\dag_K$ and $\mathcal{U}_K$ would have cancelled each other. Because the ensuing time-evolution operator commutes with $\delta\mathcal{O}_\mathrm{BRS}(t_0)$, the problem reduces to the calculation of the expectation value of the operator (\ref{eq:vector power}) at $t_0$. 

It is now obvious that for BRST-invariant states of the form (\ref{eq:BRST inv}), such an expectation value is ill-defined, because $\delta\mathcal{O}_\mathrm{BRS}(t)=\delta\mathcal{O}_\mathrm{BRS}(t_0)$ does not mix minimal and non-minimal sectors, and the BRST-invariant states
$
	|x^\pm_\mathrm{BRS}\rangle\equiv  |x^\pm, \delta b_\pm=0,
	\delta \eta^\pm=0, \delta C_\pm=0
	\rangle
$
have an ill-defined norm.  We thus select a non-zero $\delta J$ that mixes sectors, say,
\begin{subequations}\label{eq:J vector}
\begin{align}
	\delta J&=-T\sum_{\sigma=\pm 1} \int d^3p\, \delta \lambda^\sigma(\vec{p}) \delta \mathcal{P}_\sigma(\vec{p}),
	\\ 
	\mathbbm{1}_J&\equiv \exp\left(i \{\delta J, \delta^{(2)} \Omega\}\right)=
	\exp\left[T \sum_{\sigma=\pm 1}\int d^3 p\, 
	\left(i\delta \lambda^\sigma G_\sigma
	+\delta\rho^\sigma \delta\mathcal{P} _\sigma
	\right)
	\right],
\end{align}
\end{subequations}
where $T$ is an arbitrary constant with dimensions of time.  With this choice, it is a straight-forward exercise in Fourier transforms to show that
\begin{equation}\label{eq:vector expectation}
 	\langle\tilde  x{}^\pm_\mathrm{BRS}|
	\mathbbm{1}_J(t_0)
 G_+(t_0,\vec{p}_2)G_+(t_0,\vec{p}_1)
 \mathbbm{1}_J(t_0)
 | x^\pm_\mathrm{BRS}\rangle
 =0,
\end{equation}
regardless of the values of $\tilde x^\pm$ and $x{}^\pm.$  Along the same lines, the matrix element of the BRST-invariant extension of the identity operator can be  seen to equal 
\begin{equation}\label{eq:vector identity}
	\langle\tilde x{}^\pm_\mathrm{BRS}|\mathbbm{1}_J|x^\pm_\mathrm{BRS}\rangle=
	\int d(TG_\sigma) \,
	\delta[T\, G_\sigma]\exp\left(i\sum_{\sigma=\pm} \int d^3 p\,  G_\sigma(\vec{p})[\tilde x{}^\sigma(\vec{p})-x^\sigma(\vec{p})]\right),
\end{equation}
which is ill-defined for $T=0$ (it equals $0\times \infty$), but simplifies to  one for $T\neq 0$. The factor of $T$ inside the integral measure is the contribution of the fermionic sector, and the factor of $T$ inside the delta function is that of the  bosonic sector.  Note that in both equations (\ref{eq:vector expectation}) and (\ref{eq:vector identity})  the matrix element does not depend on the values of $\tilde x^\pm$ and $x^\pm$, as expected from the projected kernel.

\subsection{Quantization: Scalars}

We proceed now to illustrate Dirac and BRST quantization in the scalar sector, in which both methods are non-trivial. The fields in this sector consist of the gauge-invariant $\zeta$ and its conjugate $\Pi$ and the two gauge-variant variables $(x^L,x^N)\equiv x$  with the constraints $(G_L,G_N)\equiv G$ as their conjugates.  

\subsubsection{Dirac Quantization}
\label{sec:Free Synchronous Gauge}

In the Dirac method, one quantizes the metric in synchronous coordinates.  Although synchronous gauge was the gauge initially chosen by Lifshitz in his seminal article on cosmological perturbation theory \cite{Lifshitz:1945du}, synchronous gauge has been  widely criticized most notably  because the conditions $\delta\lambda^N=\delta\lambda^i=0$ do not fix the gauge uniquely, since  it is still possible to perform non-trivial gauge transformations that preserve the  synchronous gauge conditions \cite{Mukhanov:1990me}. To see  this, consider the transformation properties of the  linear perturbations under the linear  gauge transformations (\ref{eq:Delta delta linear}),
\begin{subequations}\label{eq:gauge transformations}
\begin{align}
	\Delta \delta \lambda^N&=a\left(\dot\xi^0+\mathcal{H}\xi^0\right),\\
	\Delta \delta \lambda_i&=\dot{\xi}_i-\partial_i \xi^0,\\
	\label{eq:Delta h}
	\Delta \delta h_{ij}&=a^2\left(2\mathcal{H}\xi^0 \delta_{ij}+\partial_i\xi_j+\partial_j \xi_i\right),\\
	\Delta \delta\varphi&=\xi^0 \, \dot{\bar\varphi},
\end{align} 
\end{subequations}
where we have set $\xi^N=a\, \xi^0$ (as mentioned above, $\delta G_N$ generates diffeomorphisms along the unit normal to the constant time hypersurfaces.)  The most general gauge transformation that preserves synchronous gauge therefore is
\begin{equation}\label{eq:gauge parameters}
	\xi^0=\frac{A(\vec{x})}{a}, \quad \xi_i=B(\vec{x})+\partial_i A \int^t \frac{d\tilde{t}}{a},
\end{equation}
which in fact is non-trivial for any non-zero choices of the free functions  $A$ and $B$. This residual gauge freedom implies that solutions to the equations of motion for the perturbations in synchronous gauge are not unique, since $\Delta \delta h_{ij}$  in (\ref{eq:Delta h}) with gauge parameters (\ref{eq:gauge parameters}) is always a solution of the synchronous linear perturbation equations. That is mostly why synchronous gauge has been essentially abandoned for analytical studies of cosmological perturbations, although it still plays a prominent role in numerical calculations.

Yet when we solve the equations of motion for the perturbations, we need to impose appropriate boundary conditions to single out a unique solution. When we extremize the action, for instance, we are typically interested in boundary conditions in which the perturbations at the endpoints are specified. Similarly,  in an initial value problem we prescribe the  values of the fields and their time derivatives at some initial time. Clearly, from equations (\ref{eq:gauge transformations}), the only choice that does not alter such  boundary conditions is $\xi^0=\xi_i=0$. In this context, the apparent gauge freedom of synchronous gauge disappears, and does not pose any particular problem. 

With this understanding in mind, let us hence consider the time evolution operator $\mathcal{U}_D$ for the perturbations (\ref{eq:U Dirac}).   The main advantage of this approach is that there are no ghosts to deal with, at any order. In addition, in synchronous gauge the action is a local functional of the perturbations.

Let us calculate the expectation of a gauge-invariant operator $\delta \mathcal{O}(t)$ such as the power spectrum of $\zeta$ in synchronous gauge. In this case, from equations (\ref{eq:non-diagonal vev all}) the expectation value is
\begin{equation}\label{eq:synchronous power}
	\langle \delta \mathcal{O}(t)\rangle=
	\int d\zeta\, dx \,
	\mathcal \psi^*_\mathrm{in}[\zeta, x]  \, 
	\delta[x]
	\langle \zeta, x | \mathcal{U}^\dag_D(t,t_0) \delta \mathcal{O}(t_0)\, \mathcal{U}_D(t,t_0) | \Psi_\mathrm{in}\rangle,
\end{equation}
where the in-state satisfies the constraints 
$
	G_L(t_0) |\Psi_\mathrm{in}\rangle=G_N(t_0) |\Psi_\mathrm{in}\rangle=0,
$
and we have chosen $f=x$ in the regularization factor $\mu=\det\{f,G\} \delta[f]$.
Because $\delta \mathcal{O}(t_0)$ commutes with  $G_{L,N}(t_0)$ by gauge-invariance,  both time evolution operators in equation (\ref{eq:synchronous power}) act on a zero eigenstate of the constraints, so we can set $G_L=G_N=0$ in the scalar Hamiltonian (\ref{eq:Hs}).  The time evolution operator reduces then to that of the gauge-invariant approach
\begin{equation}\label{eq:scalar GI}
	\mathcal{U}^\dag_D(t,t_0) \delta  \mathcal{O}(t_0) \, \mathcal{U}_D(t,t_0) |\Psi_\mathrm{in}\rangle=\mathcal{U}_\mathrm{GI}^\dag(t,t_0)\delta  \mathcal{O}(t_0) \, \mathcal{U}_\mathrm{GI}(t,t_0) |\Psi_\mathrm{in}\rangle,
\end{equation}
in which $\mathcal{U}_\mathrm{GI}$ is determined by the action of equation (\ref{eq:S zeta R}). Inserting equation (\ref{eq:scalar GI})  into equation (\ref{eq:synchronous power})  finally yields
\begin{equation}\label{eq:vev GI}
	\langle \delta \mathcal{O}(t) \rangle=\int
	  d\zeta \, d\tilde\zeta \,
	\mathcal \psi^*_\mathrm{in}[\zeta]  \, 
	\langle \zeta |
	\mathcal{U}_\mathrm{GI}^*(t, t_0)
	\delta \mathcal{O}(t_0)\, 
		\mathcal{U}_\mathrm{GI}(t, t_0)
		|\tilde\zeta\rangle \,
	 \psi_\mathrm{in}[\tilde\zeta],
\end{equation}
where we have set $\psi_\mathrm{in}[\zeta]\equiv \psi_\mathrm{in}[\zeta,x]$ (because of the constraints, the in-state wave function is $x$-independent).  Equation (\ref{eq:vev GI})  is what one would write down in the gauge-invariant formalism. Hence, from now on  the calculation follows the standard route, and expectation values of gauge-invariant operators in synchronous gauge automatically agree with those obtained in the gauge-invariant formalism. 

One of the main disadvantages of  synchronous gauge is that our scalar variables  are still coupled to each other, which renders the calculation of some of the  propagators in the scalar sector difficult.   As we describe in Appendix \ref{sec:Diagonalization of the Scalar Hamiltonian}, to decouple the scalar variables we need to carry out a canonical transformation whose coefficients contain time integrals, as opposed to local expressions in time. In the new variables, the scalar sector Hamiltonian takes the form (\ref{eq:Hs diagonal}). In the new variables, the calculation of the propagators in the scalar sector is straight-forward. Recall that  in the in-in formalism, there are four different types of propagators \cite{loops}. The four types can be constructed from appropriately ordered expectation values of  field bilinears, which, from the structure of equations (\ref{eq:non-diagonal vev all}) with $\mu=\delta[\underline x]$ require the calculation of matrix elements of the form 
\begin{equation}
	 \int d\underline\zeta\, d\tilde{\underline\zeta} \, 
	 \psi^*_\mathrm{in}[\underline\zeta]\,
		\langle \underline{\zeta},\underline{x}=0 |
		\underline{z}^i(t_2) \underline{z}^j(t_1)
		|\tilde{
		\underline\zeta},\tilde{\underline G}=0\rangle\,
	 \psi_\mathrm{in}[\tilde{\underline\zeta}],
\end{equation}
where the $z^i$ stand for any of the fields or conjugate momenta in the theory.  Therefore, quadratic matrix elements in the gauge-invariant sector spanned by $\underline{\zeta}$ and $\underline{\Pi}$ are the same as in the gauge-invariant method. Among the remaining bilinears, the only  non-vanishing matrix elements are
\begin{subequations}\label{eq:synchronous propagators}
\begin{align}
	\langle \Psi_\mathrm{in}| \underline{G}_{\sigma_2} (t_2,\vec{p}_2) \, \underline{x}^{\sigma_1} (t_1,\vec{p}_1)|\Psi_\mathrm{in}\rangle&=-i\,\delta_{\sigma_2}^{\sigma_1}\,\delta(\vec{p}_2-\vec{p}_1),\\
	\langle \Psi_\mathrm{in}| \underline{x}^{\sigma_1} (t_2,\vec{p}_2) \, \underline{x}^{\sigma_2} (t_1,\vec{p}_1)|\Psi_\mathrm{in}\rangle&=2i\,\delta^{\sigma_1 \sigma_2} \delta(\vec{p}_2+\vec{p}_1) \int_{t_0}^{t_1} dt \, \bar{A}_\sigma(t) .
\end{align}
\end{subequations}
These results  not only apply in the scalar sector, but also in the vector sector, where  the Hamiltonian already has the required diagonal structure (\ref{eq:Hv}). Although we shall not do so here,  for calculations beyond the free theory it may  be more convenient to use the inverse relations (\ref{eq:inverse relations}) to cast the propagators in terms of the original variables  $\delta h_{ij}$, $\delta\varphi$, and their canonical conjugates. Once the latter are known, one can study interactions between cosmological perturbations at any order without the need of any additional field transformations.

\subsubsection{Derivative Gauges}
\label{sec:Free Derivative Gauges}

As we have  emphasized earlier, BRST quantization allows for a much wider set of gauge choices. To illustrate the flexibility of the BRST formalism,  let us show how to dramatically simplify the structure of the Hamiltonian (\ref{eq:Hs}) by an appropriate choice of gauge-fixing fermion. First, since the gauge-fixed Hamiltonian should be quadratic and of ghost number minus one, it has to be linear in $\delta\mathcal{P}$ and $\delta C$, as in equation (\ref{eq:delta K}). Second,
because we do not intend to calculate matrix elements of the time-evolution operator between BRST-invariant states, it is not necessary for the gauge-fixing fermion $\delta K$ to  couple minimal and non-minimal sectors. One of the simplest choices that satisfies these conditions  has  $\sigma^a=0$. Choices for which $\sigma^a\neq 0$ can be used to enforce  conventional canonical gauge conditions, as we described earlier. 

By choosing  $\chi$ in equation (\ref{eq:delta K}) appropriately it is possible to remove terms in the Hamiltonian proportional to the gauge-invariant fields $G$  that are constrained to vanish in the original formulation of the theory. The freedom to alter  the evolution of the system in such a way  just captures our  expectation that only the dynamics of the gauge-invariant pairs $\zeta$ and $\Pi$ is physically relevant. In particular, picking
\begin{equation}
	\delta K=\int d^3 p\left[\left(\frac{3G_L}{4M^2 p^2 a^2}-\frac{p \, \zeta }{\mathcal{H}} -\frac{\Pi}{2M^2 a^2 p}-\frac{p \, x^N}{a } \right)\delta\mathcal{P}_L
	+\left(\frac{G_N}{2\dot{\bar\varphi}^2}-\frac{\mathcal{H}\,  \Pi}{a\dot{\bar\varphi}^2}\right)\delta \mathcal{P}_N
	\right]
\end{equation} 
we find that the gauge-fixed Hamiltonian in the scalar sector $\delta^{(2)}\! H_{0,s}+\{\delta K,\delta^{(2)}\Omega_s\}$ becomes
 \begin{equation}\label{eq:gauge fixed Hs}
\delta H^{(2)}_K=\frac{1}{2}\int d^3 p \left[
\frac{\mathcal{H}^2}{a^2 \dot{\bar\varphi}^2} \Pi^2
	 +\frac{a^2  p^2 \dot{\bar\varphi}^2}{\mathcal{H}^2}\zeta^2
	 \right].
\end{equation}
By a suitable choice of variables and gauge-fixing fermion we have thus  diagonalized the Hamiltonian.  At first sight it may appear that we took a long detour to arrive at the gauge-invariant formalism, but, in fact, the  situation here is quite different, because $\delta\lambda^{(L,N)}, x^{(L,N)}, \delta\eta^{(L,N)}, \delta C^{(L,N)}$ and their canonical conjugates remain part of the phase space.    

The  Hamiltonian  (\ref{eq:gauge fixed Hs}) does not depend quadratically on the momenta $G$ or $b$, so the theory  does not admit a Lagrangian formulation: Integrating over these momenta yields delta functionals $\delta[\dot{x}]$ and $\delta[\delta\dot{\lambda}]$, rather than quadratic terms in the velocities. In that sense, our choice of gauge-fixing fermion corresponds to a derivative gauge in which we implicitly impose the on-shell conditions $\delta\dot{\lambda}^a=0$ on the Lagrange multipliers. This property is shared by many other gauge-fixing fermions. By appropriate choices of $\chi$ and $\sigma$ we could have obtained a theory with a Lagrangian formulation  in which the $\delta\dot{\lambda}^a$ still vanish on-shell, at the expense of making the Hamiltonian slightly less simple.

We would like to calculate the expectation value of a gauge-invariant operator $\delta\mathcal{O}(t)$ along the lines of equations (\ref{eq:diagonal vev all}).  
Because the operator  $\delta\mathcal{O}(t)$ is gauge-invariant, in the free theory its BRST-invariant extension in equation (\ref{eq:observable extension}) can be taken to be the operator itself, ${\delta\mathcal{O}_\mathrm{BRS}(t_0)=\delta\mathcal{O}(t_0)}$. Then, since  by gauge choice  the Hamiltonian (\ref{eq:gauge fixed Hs})  equals that in the gauge-invariant formulation, the matrix elements of the projected operator  in equation (\ref{eq:vev extension}) factorize, thus becoming
\begin{multline}
	\langle \delta q|
		\delta\mathcal{O}_P(t)
	|\delta\tilde q\rangle
	=\langle \zeta | 
	\mathcal{U}_{GI}^\dag(t,t_0) \delta\mathcal{O}(t_0)\mathcal{U}_\mathrm{GI}(t,t_0)
	|\delta\tilde\zeta\rangle
	\\
	\times
	\langle x, \delta\eta=0, \delta b=0, \delta C=0| \mathbbm{1}_J \mathbbm{1}_J 
	|\tilde x,\delta\tilde\eta=0, \delta\tilde b=0, \delta\tilde C=0\rangle.
\end{multline}
In order for the last matrix-element to be well-defined, $\mathbbm{1}_J$ needs to couple the minimal and non-minimal sectors. We choose the analog of the gauge-fixing fermion (\ref{eq:J vector}) in the vector sector,
\begin{subequations}
\begin{align}
	J&=-T\sum_{\sigma=N, L} \int d^3p\, \delta\lambda^\sigma(\vec{p}) \delta \mathcal{P}_\sigma(\vec{p}),
	\\ 
	\mathbbm{1}_J&\equiv \exp\left(i [\delta J, \delta \Omega]\right)=
	\exp\left[T \sum_{\sigma=N,L}\int d^3 p\, 
	\left(i\delta \lambda^\sigma G_\sigma
	+\delta\rho^\sigma \delta\mathcal{P} _\sigma
	\right)
	\right],
\end{align}
\end{subequations}
which, as in equation (\ref{eq:vector identity}), implies that 
\begin{equation}
\langle x, \delta\eta=0, \delta b=0, \delta C=0| \mathbbm{1}_J \mathbbm{1}_J 
	|\delta\tilde x,\delta\tilde\eta=0, \delta\tilde b=0, \delta\tilde C=0\rangle=1.
\end{equation}
As expected,  then, the projected kernel of the operator $\delta \mathcal{O}(t)$ does not depend on the variables $ x$ and $\tilde x$, and agrees with that of the gauge-invariant method. To complete the calculation, we just need to fold the kernel of  $\delta\mathcal{O}_P(t)$ with the appropriate  wave functional of the in state, $\psi_\mathrm{in} [\zeta_i, x_i]$,  following equation (\ref{eq:diagonal vev}).  Because of gauge invariance,  the wave function is $x$-independent, so the convolution needs to be regularized by inserting factors of $\mu[x]$, say $\mu[x]=\delta[x]$. With $\psi_\mathrm{in}[\zeta]\equiv \psi_\mathrm{in}[\zeta,x=0]$ the expectation value of the operator becomes
\begin{equation}\label{eq:in in}
	\langle \Psi_\mathrm{in}| \delta \mathcal{O}(t)|\Psi_\mathrm{in} \rangle
	=\int  d\tilde\zeta\,  d\zeta\,
	\psi^*_\mathrm{in}[\tilde\zeta] \delta \mathcal{O}_P(t)[\tilde\zeta,0;\zeta, 0]\psi_\mathrm{in}[\zeta].
\end{equation}
Therefore, the BRST  returns the same expectation value as the gauge-invariant formalism, as it should. 

\subsubsection{Propagators in Derivative Gauges}
\label{sec:Propagators in Derivative Gauges}

Our  next goal  is to determine the propagators of the in-in formalism in the scalar sector, with a gauge-fixed Hamiltonian determined by equation (\ref{eq:gauge fixed Hs}).  As  in synchronous gauge, the propagators can be constructed from  
 various expectations of field bilinears. But in this case the states are BRST-invariant, and one needs to regularize the scalar products by inserting BRST-invariant extensions of the identity operator that mix minimal and non-minimal sectors, as we did above.  We shall thus calculate 
 \begin{subequations}
\begin{equation}
	 \langle \Psi_\mathrm{in}| \mathbbm{1}_J(t_0)\,   
	z^i(t_2) z^j(t_1) \mathbbm{1}_J(t_0)\, |\Psi_\mathrm{in} \rangle,
\end{equation}
\end{subequations}
in which,  because we are working in the Hamiltonian formulation,  the fields $z^i$ and $z^j$  run over the configuration variables and their conjugate momenta. 

Because the variables $\zeta$ and $\Pi$ decouple from the rest, all the field bilinears involving the latter agree with those of the gauge-invariant free theory, and, by symmetry, any mixed bilinear with a single factor of $\zeta$ or $\Pi$ vanishes. Because the conjugates $\delta b_a$ do not appear anywhere in the gauge-fixed Hamiltonian, there is no need to calculate bilinears that contain these fields.  Then, the only  remaining non-vanishing expectation values  are 
\begin{subequations}
\begin{align}
 \langle \Psi_\mathrm{in}| \mathbbm{1}_J(t_0)\,   
	\delta \lambda^{\sigma_2} (t_2) G_{\sigma_1}(t_1) \mathbbm{1}_J(t_0)\, |\Psi_\mathrm{in} \rangle
	&=\frac{i}{2T}\delta^{\sigma_2}_{\sigma_1},
	\\
	\langle \Psi_\mathrm{in}| \mathbbm{1}_J(t_0)\,   
	x^{\sigma_2} (t_2) G_{\sigma_1}(t_1) \mathbbm{1}_J(t_0)\, |\Psi_\mathrm{in} \rangle
	&=\frac{i}{2}\delta^{\sigma_2}_{\sigma_1}.
\end{align}
\end{subequations}
As in the case of synchronous gauge,  it may prove more convenient to calculate the propagators of the original fields $\delta h_{ij}$ and $\delta\varphi$ using the last expressions.  Note that because physical states carry ghost number zero, and both $\mathcal{U}_K$ and $\mathbbm{1}_J$ conserve ghost number, ghosts only appear in loop diagrams: No connected diagram  can be disconnected by cutting a single ghost line.

\subsubsection{Comparison with Other Gauges}

Leaving issues of renormalizability aside,  the reader may be wondering at this point  whether the BRST formalism has any advantages  with respect to the traditional approaches. One way to answer this question is to compare the gauge-fixed  Hamiltonian (\ref{eq:gauge fixed Hs}) with the one in the  often-employed comoving gauge, that of equation (\ref{eq:Hs}) with  $x^L\equiv x^N\equiv 0$,
\begin{equation}\label{eq:comoving Hs}
\delta H_\mathrm{com}^{(2)}= \int d^3p \, \bigg[
	\frac{\mathcal{H}^2\,  \Pi^2}{2a^2  \dot{\bar{\varphi}}^2} 
	+ \frac{p^2 a^2 \dot{\bar{\varphi}}^2\,  \zeta^2}{2\mathcal{H}^2}
	+\frac{3G_L^2}{4M^2p^2 a^2}
	+\frac{G_N^2}{2 \dot{\bar{\varphi}}^2}
	\\
	-\frac{p \,\zeta\,G_L}{\mathcal{H}}
	-\Pi\left(\frac{G_L}{2M^2 a^2 p}+\frac{\mathcal{H}G_N}{a \dot{\bar{\varphi}}^2}\right)\bigg].
\end{equation} 
By integrating over the canonical momenta in the Hamiltonian action we could   eliminate all the conjugate momenta and arrive at the Lagrangian formulation of the theory, but since our gauge-fixed Hamiltonian (\ref{eq:gauge fixed Hs}) does not admit such a formulation, we restrict ourselves to the Hamiltonians for comparison purposes. Enforcing the constraints $G_L=G_N=0$   in equation (\ref{eq:comoving Hs}) by integrating over the Lagrange multipliers  takes us back to the action (\ref{eq:S zeta R}). But the elimination of the multipliers  is not useful beyond the free theory,  because in comoving gauge the terms proportional to the multipliers (the constraints) contain quadratic and higher order  terms in the canonical variables, and it is more convenient to deal with them perturbatively. In the Lagrangian formulation one often integrates out the multipliers by replacing them by the solutions of their own equations of motion \cite{Maldacena:2002vr,Weinberg:2005vy}, but this procedure does not apply beyond  tree level,   because in the interacting theory the Lagrangian action is not quadratic in the multipliers.  Whatever the case, at this stage it is  obvious that the Hamiltonian (\ref{eq:gauge fixed Hs}) has a simpler structure than the Hamiltonian (\ref{eq:comoving Hs}). In particular, although the latter contains fewer variables, the former is diagonal.  As in many other instances, we have traded a larger number of variables for a simpler description of the theory. Although the question of simplicity  in the free theory is moot, it is important when one considers interactions, because a simpler structure of the propagators significantly eases perturbative  calculations.

\section{Interacting Theory}

We proceed now to study  interactions among cosmological perturbations.  Rather than studying these in all generality, we restrict ourselves for illustration to cubic order. The generalization to higher orders is then (conceptually) straight-forward. Even in the cubic theory there are subtleties such as operator ordering issues that we shall  gloss over. 

Our starting point is the original Hamiltonian (\ref{eq:H HD}) expanded to cubic order. Its Hamiltonian  is
\begin{equation}\label{eq:delta H 3}
	\delta^{(3)}\!H=\delta^{(3)}\!H_D+\delta\lambda^a \, \delta^{(2)}G_a,
\end{equation}
where the first-class Hamiltonian is that of the original theory (\ref{eq:H Dirac}) expanded to the same order
\begin{equation}
	\delta^{(3)}\!H_D=\bar\lambda^a(\delta G^{(2)}_a+\delta G^{(3)}_a).
\end{equation}
Note the absence of linear terms in $\delta^{(3)}\!H_D$, because we assume that the background satisfies the classical equations of motion. The constraints in the cubic theory read $\delta^{(2)}G_a=0$.  We shall  regard the expansion  to cubic order just as an approximation to the original theory, rather than as a theory on its own. Although we were able to interpret the quadratic theory literally as a gauge-invariant theory under a set of Abelian constraints,  in the cubic theory exact invariance under the appropriately truncated transformations is lost, and is replaced by invariance modulo terms of cubic order. Consider for instance the time derivative of the constraints in the cubic theory, under the time evolution generated by the cubic Hamiltonian (\ref{eq:delta H 3}),
\begin{equation}
	\frac{d\delta^{(2)}G_a}{dt}=
	\{\delta^{(2)}G_a,\delta^{(3)}H_D\}+\delta\lambda^b\{\delta^{(2)}G_a,\delta^{(2)}G_b\}+\frac{\partial\delta^{(2)}G_a}{\partial t}.
\end{equation}
As in the quadratic theory, by expanding the time derivative of the full constraints in the full theory  we obtain
\begin{equation}
	\{\delta^{(2)}G_a,\delta^{(3)}H_D\}+\delta\lambda^b\{\delta^{(2)}G_a,\delta^{(2)}G_b\}+\frac{\partial\delta^{(2)}G_a}{\partial t}=
	(\bar\lambda^b+\delta\lambda^b)C_{ab}{}^c\delta^{(2)}G_c+\mathcal{O}^{(3)}(\delta p_i,\delta q_i,\delta\lambda^a),
\end{equation}
which shows that in the cubic theory, the quadratic  constraints are preserved only up to terms of cubic order. 

\subsection{Dirac Quantization}

In Dirac quantization, the Hamiltonian of the theory is taken to be $\delta^{(3)}\!H_D$, and the constraints are imposed on the physical states,
$
	\delta^{(2)}G_a |\Psi_\mathrm{in}\rangle=0
$
(it is at this stage where factor-ordering ambiguities begin.) From equations (\ref{eq:non-diagonal vev}) and (\ref{eq:O Heisenberg}), expectation  values of gauge-invariant operators $\delta\mathcal{O}(t)$ are then determined by
\begin{equation}\label{eq:vev synchronous}
\langle \delta \mathcal{O}(t)\rangle=
	\int  dq_0  \,  d\tilde{q}_0 \, dq  \, d\tilde q\,
	\psi^*_\mathrm{in}(q_0)\,
	\mu(q_0)\,
	\mathcal{U}_D^*(q, t; q_0,t_0)
	\delta\mathcal{O}(t_0)(q,\tilde{q})\,
	\mathcal{U}_D(\tilde q, t;\tilde q_0,t_0)\,
	\psi_\mathrm{in}(\tilde q_0),
\end{equation}
where $\mathcal{U}_D$ is the time evolution operator determined by the first-class Hamiltonian  (\ref{eq:H BRST 3}).

The nature of the in-state in the presence of interactions deserves special attention. Ideally, we would like to choose the in-state to be the vacuum of the interacting theory. A well-known theorem of Gell-Mann and Low \cite{GellMann:1951rw} states that by adiabatically switching interactions off as one moves into the asymptotic past, one can recover an eigenstate of the full Hamiltonian from that of the free theory.  In particular, if the free eigenstate is chosen to be the free vacuum, Gell-Mann and Low's prescription is expected to return the vacuum of the interacting theory. Adiabatically switching off interactions in the asymptotic past amounts to time evolution  on a time contour with an infinitesimal positive imaginary component, $t\to t(1-i\epsilon)$. For our purposes, what matters is that with interactions turned off at the infinite past, we can assume the theory to be free as  $t_0\to -\infty$. In that case, our results of Section \ref{sec:Free Synchronous Gauge} apply, and we can take the wave-function $\psi_\mathrm{in}[\delta q]$ to be that of the free vacuum.  In addition, because the theory  is free in the asymptotic past, we can choose the regularization factor $\mu(\delta q)$ to be that of the free theory.  

With these choices, the calculation of the expectation value (\ref{eq:vev synchronous}) proceed as usual, either by switching to the interaction picture, or by relying on the path integral. In particular, all the propagators of the theory can be determined from the matrix elements quoted around equations (\ref{eq:synchronous propagators}). 

\subsection{Derivative Gauges}

To explore derivative gauges, we begin by expanding  the full BRST-invariant action (\ref{eq:delta S BRST non-min}) to cubic order  in the perturbations,
\begin{equation} 
	\delta^{(3)} S_K=\int dt\left[\delta p_i \delta\dot{q}^i+\delta\dot{\eta}^a\delta\mathcal{P}_a
	+\delta\dot{\lambda}^a\delta b _a+\delta\dot{C}_a\delta\rho^a-\delta^{(3)}\!H_K\right],
\end{equation}
where we have also included the contribution of the variables in the non-minimal sector.  The BRST charge here is that of equation (\ref{eq:omega nm}) expanded to the same order, 
\begin{equation}\label{eq:Omega 3}
\delta^{(3)}\Omega=\delta\eta^a \,\delta^{(2)}G_a-\frac{1}{2}\delta\eta^b \delta\eta^c \bar C_{cb}{}^a \delta\mathcal{P}_a-i\delta\rho^a \delta b_a,
\end{equation}
which corresponds to a theory in which the structure constants $\bar C_{cb}{}^a$  are non-vanishing but perturbation-independent. As  can be also seen from the BRST charge (\ref{eq:Omega 3}), the constraints are the original ones expanded to second order.
The BRST extension of the  Hamiltonian is that of equation (\ref{eq:BRST Hamiltonian})  to cubic order
\begin{equation}\label{eq:H BRST 3}
	\delta^{(3)}\!H_\mathrm{BRS}=
	\delta^{(3)}\!H_D
	-\bar{\lambda}^b \delta\eta^c (\bar{C}_{cb}{}^a+\delta C_{cb}^{(1)}{}^a)\delta\mathcal{P}_a,
\end{equation}
and the gauge-fixed Hamiltonian is constructed  from the former by adding a BRST-exact term as usual,
$
	\delta^{(3)}\!H_K=\delta^{(3)}\!H_\mathrm{BRS}+\{\delta K,\delta^{(3)}\Omega\}.
$
To preserve the cubic structure of the action $\delta K$ should be quadratic at most. In particular, we can choose the same gauge-fixing fermions as in the free theory. Note that at this stage, the Hamiltonian already contains interactions between the ghosts and the cosmological perturbations. 

We shall concentrate on the class of gauge-fixing fermions (\ref{eq:delta K}) that resulted in the free scalar Hamiltonian (\ref{eq:gauge fixed Hs}). These were characterized  by functions $\chi^a$ linear in the canonical variables, and  vanishing $\sigma^a$. With this choice, the gauge-fixed Hamiltonian becomes
\begin{equation}
	\delta^{(3)}\! H_K=\delta H^{(2)}_K+\delta H^{(3)}_\mathrm{BRS}-\chi^a \delta G_a^{(2)}+\delta\eta^a \{\delta G_a^{(2)},\chi^b\} \delta\mathcal{P}_b-\chi^a\delta\eta^b \bar C_{ab}{}^c\delta\mathcal{P}_c.
\end{equation}
Therefore, the original gauge-fixing fermion preserves the gauge-fixed Hamiltonian we derived in the free theory, $\delta H^{(2)}_K$, and introduces new interactions between the canonical variables and the ghosts.  

Our ultimate goal is to calculate expectation values of gauge-invariant operators in the interacting theory. In derivative gauges, the latter are determined by  equations (\ref{eq:diagonal vev all}).  As in synchronous gauge, the theory becomes free in  the limit $t_0\to -\infty\times(1-i\epsilon)$ so the factors of $\mu$ can be taken to be those in the free theory, $\mu=\delta[x]$. The same comments apply to the extensions $\mathbbm{1}_J(t_0)$, which can be taken to be those of the free theory. Therefore, from now on calculations  proceed perturbatively as usual: The expectation value  is split into the exponential of a quadratic piece, which includes  quadratic contributions from the $\mu$ and the $\mathbbm{1}_J$,  plus interactions, which include the cubic pieces from the Hamiltonian. By expanding the integrand in powers of these interactions one gets different moments of Gaussian integrals, which can be evaluated using the field bilinears described in  Section \ref{sec:Propagators in Derivative Gauges}. Whereas at cubic order a gauge-invariant operator is automatically BRST-invariant, at higher orders in perturbation theory one would need to replace the gauge-invariant operator $\delta\mathcal{O}$ by an appropriate BRST  extension. 

To conclude let us note that in the cubic theory neither the gauge-fixed action nor the BRST-extension of $\delta\mathcal{O}$ depend on the Lagrange multipliers $\delta\lambda^a$. In particular, quite remarkably, nothing in the action indicates that the perturbations need to satisfy the quadratic constraints $\delta^{(2)}G_a=0$, even though the in-state only obeys the linear constraints $\delta^{(1)}G_a=0$ in the asymptotic past.

\section{Summary and Conclusions}

There are basically two approaches to quantize a gauge theory: One can quantize a complete set of gauge-invariant variables in phase space, a method known as reduced phase space quantization, or one can simply fix the gauge by appropriately modifying the action of the theory.  The non-linear nature of general relativity   makes the first approach impractical, so in this article we have pursued the second approach. 

In this article we have studied the BRST quantization of cosmological perturbations in a theory with a scalar field  minimally coupled to gravity.  BRST quantization is not just yet another way of quantizing cosmological perturbations, but  is in fact necessary in any canonical gauge calculation beyond one loop. BRST quantization also offers a very general and flexible framework to quantize cosmological perturbations. For special choices of the gauge-fixing fermion in the time-evolution operator it produces the standard canonical gauge conditions of the standard quantization methods. When the gauge-fixing fermion is taken to vanish,  for appropriate state choices, it reproduces the kernel of the evolution operator in Dirac quantization, which is  the same one would write down in synchronous gauge. Finally,  when the BRST-extension of the observable mixes minimal and non-minimal sectors it allows for a wide variety of derivative gauges that cannot be reached by  other methods. We have mostly explored synchronous and derivative gauges here. 

The main advantage of Dirac quantization is the absence of  ghosts, and the relative simplification of the action implied by the vanishing of the Lagrange multipliers. Although it is often argued that the synchronous conditions implicit in Dirac quantization do not fix the gauge uniquely, this residual gauge freedom disappears when boundary conditions need to be preserved.   Even though it appears that the Dirac action does not contain information about the constraints of the theory, the latter are actually imposed on the states of the system, and the structure of the action guarantees that they are preserved by the time evolution. Actually, when interactions are adiabatically switched off in the asymptotic past, the in-state only needs to satisfy the free constraints.  The main disadvantage of Dirac quantization is that in order to fully diagonalize the Hamiltonian one needs to perform canonical transformations with coefficients that depend on the expansion history, which makes the calculation of the propagators more cumbersome. 

We have also explored derivative gauges here, in which new terms  are added or subtracted from  the first-class (or Dirac) Hamiltonian of the theory. Rather then restricting the values of the canonical variables or the Lagrange multipliers, these gauges impose \emph{on-shell} restrictions on the time derivatives of the latter.  One advantage of these derivative gauges is that by appropriate choice of the gauge-fixing fermion one can diagonalize the free Hamiltonian immediately.  The resulting simplification of the propagators then allows one  to easily carry out perturbative calculations in the interacting theory. But as in the standard approaches, the drawback of this method is that the ghosts do not decouple from the variables in the bosonic sector, and one has to include their contributions in loop calculations. 

We have shown that the structure of the Hamiltonian in derivative gauges can be much simpler than in the popular comoving gauge, even if the former contains more variables. Although the free Hamiltonian of comoving gauge dramatically simplifies when one imposes the constraints, this simplification is not particularly useful in the interacting theory,  because the constraints change with the inclusion of interactions. As in Dirac quantization, the simplified structure of the Hamiltonian in derivative gauges, combined with the theorem of Gell-Mann and Low, allows one to bypass the solution of  the constraints beyond  the linearized  theory.  

An almost unavoidable technical disadvantage of both Dirac and BRST quantization in derivative gauges is that observables are required to be gauge-invariant. Therefore, only expectation values of gauge-invariant operators have an immediate physical interpretation, and only these are guaranteed not to depend on the choice of gauge-fixing fermion.  By contrast, in canonical gauges one can always assume that any operator is the restriction of a gauge-invariant function to that gauge, so any operator can be identified with an observable. Another advantage of canonical gauges is that the antighosts $\delta\mathcal{P}_a$ are constrained to vanish in the gauge-fixed Hamiltonian, so one can ignore  that the algebra of diffeomorphisms is open. 

Our analysis has mostly focused on the operator quantization of the perturbations. The transition to the Lagrangian path integral formulation is straight-forward, as it only involves integration over the conjugate momenta of the variables in the Hamiltonian. In some of the gauges we have discussed the action is linear in the momenta, and such a Lagrangian formulation does not exist.  

Finally, we should also point out that many of  our results may be useful beyond the context of BRST quantization. We have written down for instance  the free classical Hamiltonian in a form that immediately allows one to identify the gauge-invariant variables, the constraints and their algebra. This form could be useful to shed further insights into the dynamics of cosmological perturbations in the Hamiltonian formulation and beyond. 
 
\begin{acknowledgements}
It is a pleasure to  thank Jayanth T.~Neelakanta for collaboration during the initial stages of this project,  and  Andrei Barvinsky for valuable criticism and feedback on  an earlier version of this manuscript.  A significant portion  of this work was completed while CAP was affiliated with Syracuse University. G\c{S} would like to thank Scott Watson for his kind support throughout this work. The authors would also like to thank an anonymous referee for useful comments and suggestions. 
\end{acknowledgements}

\begin{appendix}

\section{Generalized Bracket and Graded Commutator}
\label{sec:Graded Commutator}

Given two functions $F$ and $G$ of the extended phase space variables $( q^i,p_i)$ and $(\eta^a, \mathcal{P}_a)$ of definite Grassmann parity $\epsilon_F$ and $\epsilon_G$, we define their generalized Poisson bracket by
\begin{equation}
\{F,G\}=\left[\frac{\partial F}{\partial q^i}\frac{\partial G}{\partial p_i}-\frac{\partial F}{\partial p_i}\frac{\partial G}{\partial q^i}\right]
+(-)^{\epsilon_F}\left[\frac{\partial_L F}{\partial \eta^a}\frac{\partial_L G}{\partial \mathcal{P}_a}+\frac{\partial_L F}{\partial \mathcal{P}_a}\frac{\partial_L G}{\partial \eta^a}\right],
\end{equation}
where  $\partial_L/\partial$ denotes a left derivative. 
The generalized bracket obeys the algebraic properties
\begin{align}
	\{F,G\}&=-(-)^{\epsilon_F\epsilon_G} \{G,F\}
	\label{eq:P symmetry} \\
	\{F,G H\}&=\{F,G\}H+(-)^{\epsilon_F\epsilon_G}G\{F,H\} \label{eq:P Leibnitz}
\end{align}
and the Jacobi identity
\begin{equation}\label{eq:Jacobi}
	\{\{F_1,F_2\},F_3\}
	+(-)^{\epsilon_1(\epsilon_2+\epsilon_3)}
	\{\{F_2,F_3\},F_1\}
	+(-)^{\epsilon_3(\epsilon_1+\epsilon_2)}
	\{\{F_3,F_1\},F_2\}=0.
\end{equation}
When multiplied by $i$, the algebraic properties of the generalized bracket match those of the graded commutator of two operators $\hat{F}$ and $\hat{G}$,
\begin{equation}
	[\hat{F},\hat{G}]=\hat{F}\hat{G}-(-1)^{\epsilon_F \epsilon_G} \hat{G}\hat{F}.
\end{equation}
The formal analogy between the generalized bracket and the graded commutation is exploited in  canonical quantization.

\section{Irreducible Representations}
\label{sec:Irreducible Representations}

In cosmological perturbation theory it is sometimes convenient to work with perturbations that transform irreducibly under the isometries of the cosmological background: spatial rotations and translations. We thus introduce a set of seven irreducible tensors $Q_{ij}{}^\sigma(\vec{x};\vec{p})$ and $Q(\vec{x};\vec{p})$ that we use as basis elements in an expansion of arbitrary cosmological perturbations,
\begin{equation}
	\delta h_{ij}(\eta,\vec{x})=\sum_{\sigma} \int d^3 p \, Q_{ij}{}^\sigma (\vec{x};\vec{p})
	\,  \delta h_\sigma(\eta,\vec{p}), \quad
	\delta\varphi(\eta,\vec{x})=\int d^3 p\,  Q(\vec{x};\vec{p})\delta\varphi(\eta,\vec{p}).
\end{equation}
These tensors are plane waves, 
\begin{equation}
Q_{ij}{}^\sigma(\vec{x};\vec{p})\equiv  \frac{e^{i\vec{p}\cdot \vec{x}}}{(2\pi)^{3/2}} Q_{ij}{}^\sigma(\vec{p}), \quad
Q(\vec{x};\vec{p})\equiv  \frac{e^{i\vec{p}\cdot \vec{x}}}{(2\pi)^{3/2}},
\end{equation}
with  momentum-dependent components 
\begin{subequations}\label{eq:Q cov}
\begin{align}	 
	Q_{ij}{}^{L} &=2 \delta_{ij},\\
		Q_{ij}{}^{T} &=2 \left(\frac{1}{3}\delta_{ij}-\frac{p_i p_j}{p^2}\right),
		\\	
	Q_{ij}{}^{\pm1} &=-i\left(\frac{p _i}{p}\hat{\epsilon}^\pm_j+\frac{p_j}{p}\hat{\epsilon}^\pm_i\right),\\
 Q_{ij}{}^{\pm 2} &= 2\hat{\epsilon}^\pm_i\hat{\epsilon}^\pm_j.
\end{align}
\end{subequations}
Here, $\hat{\epsilon}^\pm (\vec{p})$ are two orthonormal transverse vectors with\footnote{These vectors can be taken to be $\hat{\epsilon}^\pm=R(\hat{p})\frac{1}{\sqrt{2}}(\hat{e}_x\pm i \hat{e}_y)$, where $R(\hat{p})$ is a standard rotation mapping the $z$ axis to the $\hat{p}$ direction.}
\begin{subequations}
\begin{align}
	&\vec{p}\cdot \hat{\epsilon}^\pm=0,\\
	 &\vec{p} \times \hat\epsilon^\pm =\mp \, i \,  p \,  \hat{\epsilon}^\pm.
\end{align}
\end{subequations}
Note that the polarization vectors are complex, and that $(\hat{\epsilon}^\pm)^*=\hat{\epsilon}^\mp$. Hence, it follows that $(\hat{\epsilon}^\pm)^*\cdot  \hat{\epsilon}^\pm=\hat{\epsilon}^\mp\cdot  \hat{\epsilon}^\pm=1$, but $\hat{\epsilon}^\pm\cdot  \hat{\epsilon}^\pm=(\hat{\epsilon}^\mp)^* \cdot \hat{\epsilon}^\pm=0$. The fields $\delta h_\sigma(\vec{p})$ and $\delta\varphi(\vec{p})$ are eigenvectors of spatial translations by $\vec{a}$ with eigenvalues $\exp(-i \vec{p}\cdot \vec{a})$, and spatial rotations by an angle $\theta$ around the $\vec{p}$ axis with eigenvalues $\exp(-i m \theta)$, where $m=0$ for $\delta\varphi, \delta h_L,\delta h_T$ (scalars), $m=\pm1$ for $\delta h_\pm$ (vectors) and $m=\pm 2$ for $\delta h_{\pm\pm}$ (tensors).

We similarly decompose the canonical momenta in irreducible representations, 
\begin{equation}
	\delta \pi^{ij}(\eta,\vec{x})=\sum_{\sigma} \int d^3 p \, \tilde{Q}^{ij}{}_\sigma (\vec{x};\vec{p})
	\,  \delta\pi^\sigma(\eta,\vec{p}), \quad
	\delta\pi^\varphi(\eta,\vec{x})=\int d^3 p\,  \tilde{Q}(\vec{x};\vec{p})\delta\pi^\varphi(\eta,\vec{p}),
\end{equation}
where this time the projection tensors are plane waves of opposite momentum,
\begin{equation}
	\tilde{Q}^{ij}{}_\sigma(\vec{x};\vec{p})\equiv \frac{e^{-i\vec{p}\cdot \vec{x}}}{(2\pi)^{3/2}} \tilde{Q}^{ij}{}_\sigma(\vec{p}),\quad
	\tilde{Q}(\vec{x};\vec{p})\equiv \frac{e^{-i\vec{p}\cdot \vec{x}}}{(2\pi)^{3/2}} ,
\end{equation}
 with components given by
\begin{subequations}\label{eq:Q con}
\begin{align}
		\tilde{Q}^{ij}{}_L &=\frac{1}{6} 
		\delta^{ij},\\
	\tilde{Q}^{ij}{}_T &=\frac{3}{4}
		\left(\frac{1}{3}\delta^{ij}-\frac{p^i p^j}{p^2}\right),\\
	\tilde{Q}^{ij}{}_{\pm1} &=\frac{i}{2} \left(\frac{p^i}{p}\hat{\epsilon}_\mp^j+\frac{p^j}{p}\hat{\epsilon}_\mp^i\right),\\
	\tilde{Q}^{ij}{}_{\pm2} &=\frac{1}{2}\hat{\epsilon}_\mp^i\hat{\epsilon}_\mp^j.
\end{align}
\end{subequations}
In these expressions vector and tensor indices are raised with the Euclidean metric $\delta^{ij}$.  

The projection operators (\ref{eq:Q cov}) and (\ref{eq:Q con})  satisfy the completeness relation
\begin{equation}\label{eq:completeness}
	\sum_{ij}\int d^3x\,  \tilde{Q}^{ij}{}_{\sigma_1}(\vec{x};\vec{p}_1) Q_{ij}{}^{\sigma_2}(\vec{x};\vec{p}_2)=\delta_{\sigma_1}{}^{\sigma_2} \,\delta^{(3)}(\vec{p}_1-\vec{p}_2), 
\end{equation}
which guarantee that the transition to the variables in the helicity representation is a canonical transformation. Because by definition  the field $\delta\pi^\sigma (\vec{p})$ is canonically conjugate to $\delta h_\sigma (\vec{p})$, momentum conservation demands that the field $\delta\pi^\sigma (\vec{p})$ carry the opposite momentum and helicity as $\delta h_\sigma (\vec{p})$.

Given arbitrary metric and scalar perturbations $\delta h_{ij}(\vec{x})$ and $\delta\varphi(\vec{x})$ it is straight-forward to find their components in the basis of tensors above. Because of the completeness relations (\ref{eq:completeness}), we have that
\begin{equation}
	\delta h_\sigma(\eta,\vec{p})=\int d^3 x \, 
	\delta h_{ij}(\eta,\vec{x})\tilde{Q}^{ij}{}_\sigma(\vec{x};\vec{p}) ,
	\quad
	\varphi(\eta,\vec{p})=\int d^3 x\,  \delta\varphi(\eta,\vec{x})\tilde{Q}(\vec{x};\vec{p}),
\end{equation}
and, similarly,
\begin{equation}
	\delta \pi^\sigma(\eta,\vec{p})=\int d^3 x \, \delta \pi^{ij}(\eta,\vec{x}) Q_{ij}{}^\sigma(\vec{x};\vec{p}) ,
	\quad
	\delta\pi^\varphi(\eta,\vec{p})=\int d^3 x\,  \delta\pi^\varphi(\eta,\vec{x})Q(\vec{x};\vec{p}).
\end{equation}
It is also convenient to work with the irreducible components of the spatial vectors $\delta\eta^i$ and $\delta\lambda^i$, and those of their conjugate momenta $\delta\mathcal{P}_i$  and $\delta b_i$  We thus write for instance
\begin{align}
\delta\eta^i(\eta,\vec{x})&=\int  d^3p \,
	Q^i{}_\sigma(\vec{x};\vec{p})\, \delta\eta^{\sigma}(\eta,\vec{p}),
	&\delta\eta^{\sigma}(\eta,\vec{p})&=\int  d^3x \,
	\delta\eta^i(\eta,\vec{x})
	\tilde{Q}_i{}^\sigma(\vec{x};\vec{p}),
	\\
	\delta\mathcal{P}_i(\eta,\vec{x})&=\int  d^3p \,
	\tilde{Q}_i{}^\sigma(\vec{x};\vec{p})\, \delta\mathcal{P}_{\sigma}(\eta,\vec{p}),
	&\delta\mathcal{P}_{\sigma}(\eta,\vec{p})&=\int  d^3x \,
	\delta\mathcal{P}_i(\eta,\vec{x})
	Q^i{}_\sigma(\vec{x};\vec{p}),	
\end{align}
where the components of these tensors are
\begin{subequations}
\begin{align}
	Q^i{}_L(\vec{x};\vec{p})&=-\frac{i p^i}{p}\frac{e^{i\vec{p}\cdot \vec{x}}}{(2\pi)^{3/2}}, \quad
	\tilde{Q}_i{}^L(\vec{p};\vec{x})=\frac{i p_i}{p}\frac{e^{-i\vec{p}\cdot \vec{x}}}{(2\pi)^{3/2}},\\
	Q^i{}_\pm(\vec{x};\vec{p})&=\epsilon^i_{\pm}(\vec{p})\frac{e^{i\vec{p}\cdot \vec{x}}}{(2\pi)^{3/2}}, \quad
	\tilde{Q}_i{}^\pm{}(\vec{p};\vec{x})=\epsilon_i^{\mp}(\vec{p})\frac{e^{-i\vec{p}\cdot \vec{x}}}{(2\pi)^{3/2}}.
\end{align}
\end{subequations}
We define the Fourier components of the scalars $\delta\eta^N$ and $\delta\lambda^N$, and their conjugate momenta  $\delta\mathcal{P}_N$  and $\delta b_N$ like those of $\delta\varphi$ and $\delta\pi^\varphi$. The projected linearized constraints $G_\sigma$  are defined like their ghost counterparts $\delta\mathcal{P}_\sigma$. The BRST charge $\delta\Omega^{(2)}$ is a spatial scalar, so it does not change the transformation properties of the perturbations. 

\section{Scalar Inverse Relations}

In equations (\ref{eq:new scalar variables}) we introduced a set of variables in which the Hamiltonian simplifies considerably. In  some cases it may be convenient to return to the original variables with the help of the inverse transformations
\begin{subequations}\label{eq:inverse relations}
\begin{align}
\delta h_L &=a^2\zeta+\frac{a^2 p}{3}x^L+a \mathcal{H} x^N,\\
\delta h_T &=-a^2 p\, x^L,\\
\delta\varphi &=\frac{\dot{\bar\varphi}}{a}x^N,\\
\delta\pi_L&=\frac{2M^2 (p^2-3\mathcal{H}^2)}{\mathcal{H}}\zeta-2M^2 p \,\mathcal{H}\, x^L+\frac{3\dot{\bar\varphi}^2+2M^2(p^2-3\mathcal{H}^2)}{a}x^N+\frac{\Pi}{a^2},\\
\delta\pi_T &=\frac{2M^2 p^2}{3\mathcal{H}}\zeta-\frac{8 M^2 p\, \mathcal{H}}{3}x^L+\frac{2M^2 p^2}{3a}x^N+\frac{\Pi}{3a^2}-\frac{G_L}{a^2 p},\\
\delta\pi_\varphi&=3a^2 \dot{\bar\varphi}\,\zeta+a^2 p \,\dot{\bar\varphi}\,x^L-a^3 \bar{V}_{,\varphi}\, x^N-\frac{\mathcal{H}}{\dot{\bar\varphi}}\Pi+\frac{a}{\bar{\dot\varphi}}G_N.
\end{align}
\end{subequations}
We have suppressed here the momentum arguments for simplicity. The latter can be restored by noting that conjugate momenta have momenta opposite to those of their conjugates.

\section{Diagonalization of the Scalar Hamiltonian}
\label{sec:Diagonalization of the Scalar Hamiltonian}

By an appropriate canonical transformation it is possible to further simplify the structure of the Hamiltonian (\ref{eq:Hs}). Rather than performing such a canonical transformation at once, it is more convenient to carry out a chain of transformations, each one chosen to eliminate a targeted set of non-diagonal terms. The first canonical transformation  eliminates the mixed term $x^NG_L$ in the Hamiltonian. 

\begin{subequations}\label{eq:canonical 1}
\begin{align}
	x^L&\to x^L-\alpha_G\,x^N+\beta_G\, G_N,
	\\
	x^N&\to x^N+\beta_G\,G_L,
	\\
G_N&\to G_N+\alpha_G\, G_L,
\end{align} 
\end{subequations}
where 
\begin{align}
\alpha_G&=\int ^t \frac{p}{a}\,
d\tilde{t},
\\
\beta_G&=\int^t \frac{\alpha_G}{\dot{\bar{\varphi}}^2}  \, d\tilde{t}.
\end{align}
Note that in order to find the transformed Hamiltonian it is not necessary to calculate the generating function. Instead, one simply substitutes the transformation (\ref{eq:canonical 1}) into the action and reads off the transformed Hamiltonian. 

In order to remove the coupling between  $\Pi$ and $G_N$, while keeping the term we just eliminated absent, we redefine
\begin{subequations}\label{eq:canonical 2}
\begin{align}
\zeta&\to\zeta+\alpha_N\, G_N, \\
x^L & \to x^L+\beta_N \, G_N, \\
x^N &\to x^N-\gamma_N\,  \zeta+ \alpha_N \, \Pi+\beta_N \,G_L, \\
\Pi&\to\Pi +\gamma_N \,G_N,
\end{align}
where $\alpha_N$, $\beta_N$  and $\gamma_N$ satisfy
\begin{align}
	\ddot{\alpha}_N &=
	\left(4H+\frac{2a^2 \bar{V}_{,\varphi}}{\dot{\bar{\varphi}}}-\frac{\dot{\bar{\varphi}}^2}{M^2 \mathcal{H}}\right)
	\dot\alpha_N
	-p^2 \alpha_N
	-\frac{1}{2M^2a}, \label{eq:Dalpha2}
	\\
	\beta_N&=-\int^t \left(
	\frac{1}{2M^2 p\,  a \mathcal{H}}+\frac{\alpha_G}{\dot{\bar{\varphi}}^2}
	+\frac{p\,\alpha_N}{\mathcal{H}}
	+\frac{\dot{\bar{\varphi}}^2\,\dot{\alpha}_N}{2M^2 p \,\mathcal{H}^2}+\frac{a\, \alpha_G\,\dot{\alpha}_N}{\mathcal{H}}\right) d\tilde t,
	\\
	\gamma_N&=\frac{a}{\mathcal{H}^2}\left(\mathcal{H}+a\, \dot{\bar{\varphi}}^2 \dot{\alpha}_N\right).
\end{align}
\end{subequations}

Finally, to get rid of the mixing between the $\zeta$ and $x^L$ sectors we introduce
\begin{subequations}\label{eq:canonical 3}
\begin{align}
\zeta&\to\zeta+\beta_L \, G_L, \\
x^L &\to x^L-\alpha_L \, \zeta+ \beta_L \, \Pi+\gamma_L\,G_L, \\
\Pi&\to\Pi +\alpha_L \,G_L,
\end{align}
\end{subequations}
where the functions $\alpha_L, \beta_L$ and $\gamma_L$ obey
\begin{align}
	\ddot{\alpha}_L&=\left(\frac{\dot{\bar{\varphi}}^2}{M^2 \mathcal{H}}-4\mathcal{H}-\frac{2a^2 \bar{V}_{,\varphi}}{\dot{\bar{\varphi}}}\right)
	\dot\alpha_L
	-p^2\alpha_L
	+\frac{p^2 a\,\alpha_G}{\mathcal{H}}
	+3p+\frac{2p\, a^2\, \bar{V}_{,\varphi}}{\dot{\bar\varphi}\mathcal{H}},
	\\
\beta_L&=\frac{\mathcal{H} \left(p-\mathcal{H}\dot\alpha_L\right)}{p^2\, a^2\, \dot{\bar\varphi}^2},
\\
	\gamma_L&=-p \int^t \frac{\beta_G}{a} d\tilde{t}.
\end{align}
The end-result of the successive canonical  transformations (\ref{eq:canonical 1}), (\ref{eq:canonical 2}) and (\ref{eq:canonical 3})  is a Hamiltonian that in the final variables, denoted by a tilde, has the form of equation (\ref{eq:Hs diagonal}), where the time-dependent coefficients $\bar{A}_L$ and $\bar{A}_N$ are
\begin{equation}\label{eq:As}
\bar{A}_L= \frac{(3-p\, \alpha_L)\dot{\bar\varphi}^2-2M^2 p\, (p+p\, a\, \mathcal{H}\, \alpha_G\,  \alpha_L-\mathcal{H}\,\dot\alpha_L)}{4M^2 p^2 a^2 \dot{\bar\varphi}^2},
\quad
\bar{A}_N=-\frac{a\, \dot\alpha_N}{2\mathcal{H}}.
\end{equation}
To bring the Hamiltonian to this form, we need to solve essentially two decoupled linear inhomogeneous second order differential equations, and a set of first integrals. During power-law inflation, the differential equations admit closed solutions in terms of Bessel functions.
\end{appendix}

\end{document}